\documentclass[10pt,a4paper]{article}
\usepackage[T1]{fontenc}
\usepackage{microtype}
\usepackage{amsmath,amssymb}
\usepackage{amsfonts}
\usepackage{mathrsfs}
\usepackage[table]{xcolor}  
\usepackage{booktabs}
\usepackage{array}
\usepackage{multirow}
\usepackage{longtable}
\usepackage{tabularx}
\usepackage{xltabular}

\usepackage{graphicx}
\usepackage[export]{adjustbox}
\usepackage{pdflscape}
\usepackage{rotating}
\usepackage{caption}
\usepackage{subcaption}
\usepackage{authblk}
\usepackage{etoolbox}
\usepackage{orcidlink}
\usepackage{textcomp}
\usepackage[margin=2.5cm,top=3cm]{geometry}
\usepackage{hyperref}
\usepackage[numbers,sort&compress]{natbib}
\hypersetup{colorlinks=true, linkcolor=blue, citecolor=blue, urlcolor=blue}
\providecommand{\keywords}[1]{%
  \par\vspace{0.5ex}%
  \noindent\textbf{Keywords: }#1\par
}
\numberwithin{equation}{section}
\numberwithin{figure}{section}
\numberwithin{table}{section}
\newcolumntype{C}{>{\centering\arraybackslash}X}
\newcolumntype{L}{>{\raggedright\arraybackslash}X}
\newcolumntype{R}{>{\raggedleft\arraybackslash}X}
\renewcommand{\arraystretch}{1.3}
\begin{document}

\begin{center}
\LARGE{Repetitive Penrose process for charged particles in Kerr–Newman black holes \\\;\\\;\\}
\par\end{center}

\begin{center}
{\bf Mohammad Reza Alipour\orcidlink{0000-0001-8074-7865}}\footnote{\bf mohamad.alipour.1994@gmail.com; mr.alipour@stu.umz.ac.ir}\\
{\it School of Physics, Damghan University, P.~O.~Box 3671641167, Damghan, Iran}
\end{center}

\begin{center}
{\bf Saeed Noori Gashti\orcidlink{0000-0001-7844-2640}}\footnote{\bf sn.gashti@du.ac.ir; saeed.noorigashti70@gmail.com}\\
{\it School of Physics, Damghan University, P.~O.~Box 3671641167, Damghan, Iran}
\end{center}

\begin{center}
{\bf Mohammad Ali S. Afshar\orcidlink{0009-0001-3133-5992}}\footnote{\bf m.a.s.afshar@gmail.com}\\
{\it School of Physics, Damghan University, P.~O.~Box 3671641167, Damghan, Iran}
\end{center}

\begin{abstract}
We investigate the repetitive Penrose process for charged particles in an initially
extremal Kerr--Newman black hole and develop a nonlinear iterative framework in which
the black-hole mass, angular momentum, electric charge, and irreducible mass are updated
after every extraction event. By imposing the triple turning-point condition, we obtain
an analytic solution of the conservation equations for energy, angular momentum, charge,
and radial momentum, allowing the entire extraction sequence to be followed
self-consistently.
The analysis shows that the dynamics are controlled by two independent electromagnetic
couplings. The coupling $\hat Q\hat q_0$ determines whether the incident particle can
continue to access the ergoregion and therefore governs the termination of the repetitive
process, whereas the coupling $\hat Q\hat q_1$ controls the depth of the negative-energy
states and consequently the extraction efficiency. An attractive interaction between the
black hole and the captured fragment ($\hat Q\hat q_1<0$) substantially enhances both
the energy return on investment and the energy utilization efficiency and, above a
critical charge, produces a transient increase of the dimensionless spin parameter even
though the total angular momentum decreases.
We further identify a four-region structure in the captured-particle charge parameter
space, consisting of two physically allowed regions separated by a forbidden interval and
terminated by a thermodynamically forbidden regime. Near the critical charge
$\hat q_1\simeq-54.85405$, the evolution approaches the reversible
Christodoulou--Ruffini limit, with the energy utilization efficiency approaching unity
while the black hole remains sub-extremal. Beyond this point the irreducible mass
decreases, indicating the breakdown of the test-particle approximation.
Unlike the repetitive Penrose process in the extremal Reissner--Nordström spacetime, the
Kerr--Newman black hole can evolve continuously through the neutral state and reverse the
sign of its electric charge without violating the area theorem or cosmic censorship.
These results show that allowing the black hole to evolve simultaneously through its
rotational and electromagnetic degrees of freedom qualitatively changes the extraction
process. In particular, the electric charge is no longer prevented from passing through
the neutral state, demonstrating that the discharge barrier found in the
Reissner--Nordstr\"om case is a consequence of the absence of rotation rather than a
general property of charged black holes.
\end{abstract}


\tableofcontents

\section{Introduction}

The extraction of energy from black holes has been a central theme of relativistic
astrophysics since Penrose demonstrated that the rotational energy of a Kerr black hole
can, in principle, be tapped by a particle that decays within its ergosphere
\cite{Penrose1969,PenroseFloyd1971}. The mechanism rests on the existence of negative-energy orbits in
the ergoregion, where the asymptotic time-translation Killing vector becomes spacelike:
an incident particle splits into a fragment that crosses the horizon carrying negative
energy and a fragment that escapes to infinity with more energy than the parent. The
detailed energetics of this mechanism, including the effective-potential description of
equatorial orbits and the resulting bound on the maximum extractable rotational energy,
were subsequently established by Bardeen, Press, and Teukolsky and by
Wald~\cite{Bardeen1972,Wald1974}.
Soon afterward, Denardo and Ruffini showed that an analogous ``generalized ergosphere''
and negative-energy states arise for charged particles in the Reissner--Nordstr\"om
geometry, giving rise to the electric Penrose process and establishing that
electromagnetic interactions can drive energy extraction in the absence of rotation
\cite{Denardo1973,Ruffini1975}. These ideas have since matured into a broad family of
extraction mechanisms, including superradiance, the Blandford--Znajek process, and
magnetic and magnetohydrodynamic variants of the Penrose mechanism.

A decisive conceptual refinement was introduced only recently. Ruffini and collaborators
\cite{Ruffini2025PRL,Ruffini2025PRR} revisited the long-standing proposal of a
\emph{repetitive} Penrose process---in which decays are repeated, one after another,
to drain the rotational energy of the hole---and showed that the naive linear
implementation, in which the black-hole mass and spin are held fixed between events,
violates energy conservation. Once each iteration is treated self-consistently, by
updating the mass and angular momentum and accounting for the irreducible mass
$M_{\mathrm{irr}}$ \cite{Christodoulou1970,ChristodoulouRuffini1971,Hawking1971}, the
process becomes intrinsically nonlinear. The key result is that the bulk of the lost
extractable energy is converted not into escaping radiation but into a monotonic growth
of $M_{\mathrm{irr}}$---and hence of the horizon area, in accordance with Hawking's area
theorem, the classical analogue of the second law of black-hole
thermodynamics~\cite{Bekenstein1973}. As a consequence the sequence terminates after a finite number of steps at a
positive lower bound on the spin, and the total extracted energy is only a small fraction
of the black-hole mass: the full rotational energy of an extremal Kerr black hole can
never be classically exhausted.

This framework has been rapidly extended to richer spacetimes. Hu, Cai, and Wang
\cite{HuCaiWang2026} formulated the repetitive \emph{electric} Penrose process in
Reissner--Nordstr\"om black holes and found a closely analogous obstruction: the charge
cannot be reduced exactly to zero through a finite sequence of classical decays,
which they interpreted as a thermodynamic third-law analog. Their analysis also
introduced two complementary efficiency measures---the energy return on investment
(EROI) and the energy utilization efficiency (EUE)---and established that, although the
EROI may exceed unity, the EUE rarely surpasses $50\%$, since a large fraction of the
reduced extractable energy is irreversibly locked into $M_{\mathrm{irr}}$. Wang and Zeng
\cite{WangZeng2025} subsequently examined Kerr--de~Sitter black holes and showed that a
positive cosmological constant raises both the EROI and the single-extraction capability
relative to the Kerr case, with the efficiency depending sensitively on the decay radius.
Most recently, Zeng and Wang \cite{ZengWang2026} studied accelerating Kerr black holes
described by the C-metric and reported that, at sufficiently small decay radii, the
acceleration parameter can push the EUE above $50\%$---the first setting in which the
reduced extractable energy is channeled predominantly into escaping energy rather than
into irreducible mass.

Despite this progress, the repetitive Penrose process has not yet been analyzed in the
most general stationary and asymptotically flat solution of the Einstein--Maxwell
equations---the Kerr--Newman geometry---and, in particular, for an incident particle that
itself carries electric charge. This case is qualitatively distinct from those studied so
far. Whereas the Kerr analyses involve only frame dragging and the Reissner--Nordstr\"om
analyses only electrostatic forces, the charged Kerr--Newman problem couples the two:
the motion of each fragment is governed simultaneously by gravitational attraction,
frame dragging, and the electromagnetic interaction between the particle charges
$q_i$ and the black-hole charge $Q$. This coupling reshapes the effective potentials,
modifies the negative-energy states in the ergoregion, and allows energy and charge to be
extracted in the same iteration, so that the evolution of the spin, charge, and
irreducible mass becomes intricately interdependent.

In this work we develop the repetitive Penrose process for a charged test particle in a
Kerr--Newman background. Imposing the turning-point condition on all three particles, we
obtain the complete analytic solution of the conservation laws for energy, charge, angular
momentum, and radial momentum, and construct a nonlinear iterative scheme that updates the
black-hole mass, angular momentum, and electric charge after each extraction event while
tracking the evolution of the irreducible mass $M_{\mathrm{irr}}$. We identify the
dimensionless coupling $\hat{Q}\hat{q}_0$ as the quantity controlling the competition
between gravitational attraction and electromagnetic repulsion, and show that it separates
two distinct dynamical regimes characterized by different termination conditions for the
iterative process. We further demonstrate that the charge of the captured fragment acts as
a powerful control parameter: an attractive black-hole--particle interaction
($\hat{Q}\hat{q}_1<0$) deepens the negative-energy channel, substantially enhances both
the energy return on investment (EROI) and the energy utilization efficiency (EUE),
induces a transient spin-up of the black hole, and, near a critical charge, drives the
process toward the reversible Christodoulou limit, where the EUE exceeds the
$50\%$ bound previously found for neutral Kerr and Reissner--Nordstr\"om black holes.

An unexpected result of the present analysis concerns the evolution of the black-hole
electric charge. In contrast to the repetitive Penrose process in the extremal
Reissner--Nordstr\"om spacetime, where the charge cannot be completely extracted and
remains bounded away from zero---a behavior interpreted as a classical analogue of the
third law of black-hole thermodynamics~\cite{HuCaiWang2026}---we find that, in the
extremal Kerr--Newman spacetime, the simultaneous extraction of rotational and
electromagnetic energy allows the horizon charge to evolve continuously through
$\hat{Q}=0$ and even reverse its sign while preserving both the Hawking area theorem and
the cosmic censorship condition $\hat{a}^2+\hat{Q}^2\le1$. This qualitative difference
demonstrates that the coupling between rotation and charge fundamentally modifies the
thermodynamic evolution of charged black holes. It indicates that the third-law-like
charge barrier identified for purely electric black holes is not universal, but is instead
a special property of the Reissner--Nordstr\"om geometry. Finally, we show that beyond a
second critical value of the captured-particle charge the irreducible mass would decrease,
signaling the breakdown of the test-particle approximation and the onset of self-force
and backreaction effects.

The paper is organized as follows.
Section~\ref{sec:geometry} presents the theoretical formulation of the repetitive
Penrose process in Kerr--Newman spacetime.
Section~\ref{sec:repetitive} describes the numerical implementation of the nonlinear
iterative scheme.
Section~\ref{sec:results} presents the numerical results and their physical
interpretation.
Finally, Sec.~\ref{sec:conclusion} summarizes our main findings. 
Throughout this paper we use geometrized units, $G=c=1$, with metric signature
$(-,+,+,+)$, and set $M_0=1$ in the numerical calculations.

\section{Kerr--Newman Geometry and Charged Particle Motion}\label{sec:geometry}

In this section, we briefly review the Kerr--Newman spacetime and derive the equations governing the motion of charged test particles.

\subsection{Kerr--Newman Metric}

The Kerr--Newman metric describes the most general stationary, axisymmetric, and asymptotically flat black hole solution of the Einstein--Maxwell equations, characterized by mass $M$, angular momentum parameter $a = L/M$, and electric charge $Q$. In Boyer--Lindquist coordinates $(t,r,\theta,\phi)$, the line element is
\begin{equation}
ds^2 = -\left(1 - \frac{2Mr - Q^2}{\Sigma}\right)dt^2 - \frac{2a(2Mr - Q^2)\sin^2\theta}{\Sigma}\,dt\,d\phi + \frac{\Sigma}{\Delta}\,dr^2 + \Sigma\,d\theta^2 
+ \frac{(r^2 + a^2)^2 - a^2\Delta\sin^2\theta}{\Sigma}\sin^2\theta\,d\phi^2,
\end{equation}
where
\begin{equation}
\Sigma = r^2 + a^2\cos^2\theta, \qquad \Delta = r^2 - 2Mr + a^2 + Q^2.
\end{equation}

The electromagnetic four-potential associated with the black hole charge is
\begin{equation}
A_\mu = \left( -\frac{Qr}{\Sigma},\ 0,\ 0,\ \frac{aQr\sin^2\theta}{\Sigma} \right).
\end{equation}

The event horizons of the Kerr--Newman spacetime are determined by the roots of $\Delta = 0$, which leads to the following radial coordinates:
\begin{equation}
r_\pm = M \pm \sqrt{M^2 - a^2 - Q^2},
\end{equation}
where $r_+$ and $r_-$ denote the outer (event) and inner (Cauchy) horizons, respectively. For a physically viable black hole, the quantity under the square root must be non-negative, which imposes the inequality $a^2 + Q^2 \le M^2$. The limiting case $a^2 + Q^2 = M^2$ defines the extremal black hole, for which the two horizons coincide ($r_+ = r_-$). In the extremal limit, the black hole possesses the maximum possible angular momentum and charge for a given mass, a configuration that maximizes the extractable energy and is therefore the natural starting point for our analysis of the repetitive Penrose process.

The ergoregion, which is essential for the Penrose mechanism, is bounded by the stationary limit surface, where the norm of the timelike Killing vector $\partial_t$ vanishes, i.e., $g_{tt} = 0$.
On the equatorial plane ($\theta=\pi/2$), this condition yields
\begin{equation}
r_{\rm ergo}=M+\sqrt{M^2-Q^2},
\label{eq:rergo}
\end{equation}
which reduces to $r_{\rm ergo}=2M$ in the Kerr limit ($Q=0$).
Note that Eq.~\eqref{eq:rergo} is independent of the spin parameter $a$---a
property unique to the equatorial plane in the Kerr--Newman family---while the
explicit $Q$-dependence means that horizon charge \emph{shrinks} the ergoregion,
reducing the phase space available for energy extraction.
The ergoregion is defined by the interval
\begin{equation}
r_+<r<r_{\rm ergo},
\end{equation}
Within this region, the Killing vector $\partial_t$ becomes spacelike, allowing the existence of negative-energy states as measured by asymptotic observers. This property provides the physical basis for the Penrose process: an incident particle entering the ergoregion may undergo a splitting event into two fragments, one of which follows a negative-energy trajectory and falls into the black hole, while the other escapes to infinity carrying an energy greater than that of the original particle.
Consequently, any physically viable Penrose process must occur within the ergoregion. For $r<r_+$, the black-hole interior is causally disconnected from infinity, whereas for $r>r_{\rm ergo}$, negative-energy states are forbidden and classical Penrose energy extraction becomes impossible.

The surface area of the outer event horizon of a Kerr--Newman black hole is
\begin{equation}
S = 4\pi \left( r_+^2 + a^2 \right).
\end{equation}
The irreducible mass $M_{\rm irr}$, which is directly related to the horizon area via $S = 16\pi M_{\rm irr}^2$, is given by \cite{ChristodoulouRuffini1971}
\begin{equation}
M_{\rm irr}^2 = \frac{1}{2}\left(M^2 - \frac{Q^2}{2}
+ M\sqrt{M^2 - a^2 - Q^2}\right)
= \frac{r_+^2 + a^2}{4},
\end{equation}
where the second equality makes explicit that $M_{\rm irr}$ depends only on the outer-horizon geometry.
The extractable energy of the black hole is then defined as the difference between its total mass and the irreducible mass:
\begin{equation}
E_{\rm extractable} = M - M_{\rm irr}.
\end{equation}
This quantity plays a central role in the repetitive Penrose process, as any irreversible energy extraction increases $M_{\rm irr}$ according to the area theorem, limiting the total energy that can be classically extracted.

\subsection{Fundamental Equations of the Penrose Process}

The Penrose process relies on the conservation of energy, charge, angular momentum, and radial momentum during the decay of the incident particle inside the ergoregion. Let particle 0 be the incident particle with rest mass $\mu_0$, energy $E_0$, charge $q_0$, and four-momentum $p_0^\mu$. It decays into two fragments: particle 1 (captured by the black hole with negative energy) and particle 2 (escapes to infinity with positive energy).

The conservation equations are
\begin{subequations}
\begin{align}
\hat{E}_0 &= \tilde{\mu}_1 \hat{E}_1 + \tilde{\mu}_2 \hat{E}_2, \label{eq:energy} \\
\hat{q}_0 &= \tilde{\mu}_1 \hat{q}_1 + \tilde{\mu}_2 \hat{q}_2, \label{eq:charge} \\
\hat{p}_{\phi 0} &= \tilde{\mu}_1 \hat{p}_{\phi 1} + \tilde{\mu}_2 \hat{p}_{\phi 2}, \label{eq:angmom} \\
\hat{p}_{r 0} &= \tilde{\mu}_1 \hat{p}_{r 1} + \tilde{\mu}_2 \hat{p}_{r 2}, \label{eq:radmom}
\end{align}
\end{subequations}
where the dimensionless quantities are defined as
\begin{equation}\label{eq:dimless}
\begin{split}
\hat{E}_i = \frac{E_i}{\mu_i}, \quad 
\hat{q}_i = \frac{q_i}{\mu_i}, \quad 
\hat{p}_{\phi i} = \frac{p_{\phi i}}{\mu_i M}, \quad 
\hat{p}_{r i} = \frac{p_{r i}}{\mu_i}, \quad 
\tilde{\mu}_i = \frac{\mu_i}{\mu_0}.  
\end{split}
\end{equation}

Here, $\mu_i$ and $q_i$ denote the rest mass and charge of particle $i$, respectively. The normalization condition for the canonical (generalized) four-momentum $\pi_\mu = p_\mu - q A_\mu$ reads
\begin{equation}
g^{\mu\nu}(p_\mu - q A_\mu)(p_\nu - q A_\nu) = -\mu^2,
\end{equation}
with the sign convention $(-,+,+,+)$ so that $g^{\mu\nu}p_\mu p_\nu = -\mu^2$ for an uncharged particle.

For maximum energy extraction efficiency, we impose the optimal turning-point condition, i.e., the radial momenta of all three particles vanish at the decay radius ($p_{r0} = p_{r1} = p_{r2} = 0$). This condition ensures that the decay occurs at the points where the effective potentials for particles 0, 1, and 2 reach their respective values, maximizing the negative energy carried by particle 1.

\subsection{Effective Potential and Turning-Point Condition}

For motion on the equatorial plane ($\theta = \pi/2$), we have $p_\theta = 0$. When the radial momentum is nonzero ($p_r \neq 0$), the normalization condition of the generalized four-momentum takes the form
\begin{equation}
g^{tt} \left( \hat{E}_i - \frac{\hat{Q} \hat{q}_i}{\hat{r}} \right)^2 
- 2 g^{t\phi} M \left( \hat{p}_{\phi i} - \frac{\hat{a} \hat{Q} \hat{q}_i}{\hat{r}} \right) 
\left( \hat{E}_i - \frac{\hat{Q} \hat{q}_i}{\hat{r}} \right) 
+ g^{\phi\phi} M^2 \left( \hat{p}_{\phi i} - \frac{\hat{a} \hat{Q} \hat{q}_i}{\hat{r}} \right)^2 
+ g^{rr} \hat{p}_{r i}^2 = -1.
\label{eq:norm-full}
\end{equation}

Solving this equation for the radial momentum yields
\begin{equation}
\hat{p}_{r i}^2 = -\frac{g^{tt}}{g^{rr}} 
\left( \hat{E}_i - \hat{V}_i^{+} \right) 
\left( \hat{E}_i - \hat{V}_i^{-} \right),
\label{eq:pr}
\end{equation}
where the effective potentials $\hat{V}_i^{\pm}$ are given by
\begin{equation}
\hat{{V}}_i^{\pm}=
\frac{\hat{Q} \hat{q}_i}{\hat{r}}+ \frac{M g^{\text{t$\phi $}} \left(\hat{p}_{\text{$\phi $i}}-\frac{\hat{a} \hat{Q} \hat{q}_i}{\hat{r}}\right)\mp\sqrt{\left(M g^{\text{t$\phi $}}\right)^2 \left(\hat{p}_{\text{$\phi $i}}-\frac{\hat{a} \hat{Q} \hat{q}_i}{\hat{r}}\right){}^2-g^{\text{tt}} \left(M^2 g^{\phi \phi } \left(\hat{p}_{\text{$\phi $i}}-\frac{\hat{a} \hat{Q} \hat{q}_i}{\hat{r}}\right){}^2+1\right)}}{g^{\text{tt}}}
\label{eq:Vefff}
\end{equation}

Here, the electromagnetic coupling between the particle charge $q_i$ and the black hole charge $Q$ is explicitly incorporated through the shifted combinations of energy and angular momentum. The $\pm$ branches correspond to the two possible solutions of the quadratic equation. For physical future-directed trajectories satisfying the timelike geodesic condition, the appropriate branch must be selected such that $\frac{dt}{d\tau} > 0$. This general expression reduces to the standard uncharged case when $\hat{q}_i = 0$ and recovers the turning-point potentials when $\hat{p}_{r i} = 0$.

Indeed, the above relation holds for general stationary and axisymmetric metrics. Since $g^{tt} < 0$ and $g^{rr} > 0$, the condition $\hat{p}_{ri}^2 \ge 0$ implies that $\hat{E}_i \ge \hat{V}_i^{+}$ or $\hat{E}_i \le \hat{V}_i^{-}$. 

Furthermore, physical trajectories must be future-directed and timelike, which requires $\frac{dt}{d\tau} > 0$. The coordinate time derivative is
\begin{equation}
\frac{dt}{d\tau}=  - g^{\text{tt}}\hat{E}+g^{\text{t$\phi $}} M \hat{p}_{\phi }+\frac{\hat{q} \hat{Q}}{\hat{r}} \bigg( g^{\text{tt}}-\hat{a} M g^{\text{t$\phi $}}  \bigg)
\label{eq:dt-dtau}
\end{equation}

Evaluating Eq.~\eqref{eq:dt-dtau} at $\hat{E}_i = \hat{V}_i^{\pm}$ gives
\begin{equation}
\left. \frac{dt}{d\tau} \right|_{\hat{E}_i = \hat{V}_i^{\pm}}=
\pm \sqrt{\left(M g^{t\phi}\right)^2 \left(\hat{p}_{\phi }-\frac{\hat{a} \hat{q} \hat{Q}}{\hat{r}}\right)^2-g^{tt} \left(M^2 g^{\phi \phi } \left(\hat{p}_{\phi }-\frac{\hat{a} \hat{q} \hat{Q}}{\hat{r}}\right)^2+1\right)}.
\end{equation}
The expression under the square root is strictly positive. However, when $\hat{E}_i = \hat{V}_i^{-}$, we have $\frac{dt}{d\tau} < 0$, which violates the future-directed timelike condition. Therefore, only the branch $\hat{E}_i \ge \hat{V}_i^{+}$ is physically acceptable.
For the Penrose process to extract energy, the captured particle must satisfy $\hat{E}_1 < 0$. To maximize the extracted energy, $\hat{E}_1$ should be as negative as possible. Combined with the condition $\hat{E}_1 \ge \hat{V}_1^{+}$, the most negative value occurs when particle 1 is at its turning point, i.e., $\hat{E}_1 = \hat{V}_1^{+}$. At this point, the radial momentum conservation implies $\hat{p}_{r0} = \hat{p}_{r2}$. 

Assume, for the sake of contradiction, that $\hat{p}_{r0} = \hat{p}_{r2} \neq 0$. Substituting this assumption into the radial momentum equation and applying the conservation laws for energy and angular momentum leads to a specific value for $\hat{E}_1$. 

This resulting expression for $\hat{E}_1$ violates the timelike geodesic condition $\frac{dt}{d\tau} > 0$ for the captured particle. Such a violation is physically unacceptable, as it would imply that particle 1 travels backward in coordinate time as measured by a distant observer.

Therefore, the assumption $\hat{p}_{r0} = \hat{p}_{r2} \neq 0$ must be rejected. The only configuration consistent with both radial momentum conservation and the timelike geodesic requirement is
$\hat{p}_{r0} = \hat{p}_{r1} = \hat{p}_{r2} = 0$.
Hence, at the decay point, all three particles must be located at their respective turning points. This condition maximizes the negative energy carried by the captured particle ($\hat{E}_1 = \hat{V}_1^{+}$) and provides the optimal setup for energy extraction in the Penrose process.

Under the optimal turning-point condition ($\hat{p}_{r0}=\hat{p}_{r1}=\hat{p}_{r2}=0$) at the decay radius, and assuming $\hat{E}_0$, $\hat{p}_{\phi 1}$, and the mass ratio $\nu=\mu_2/\mu_1$ are given, the fundamental conservation equations of the Penrose process in the Kerr--Newman spacetime admit the following analytic solution:

\begin{equation}
\begin{split}
\hat{p}_{\phi 0}
&= \frac{\hat{a} \hat{q}_0 \hat{Q}}{\hat{r}}
- \frac{\hat{q}_0 \hat{Q} \left(M g^{t\phi}\right)}{\hat{r} \left(M^2 g^{\phi\phi}\right)}
+ \frac{
    \hat{E}_0 \left(M g^{t\phi}\right) + \sqrt{R}
}{
    M^2 g^{\phi\phi}
}, \\
R ={} &-M^2 g^{\phi\phi}
-\frac{2 \hat{E}_0 \left(\hat{q}_0 \hat{Q}\right) \left(M g^{t\phi}\right)^2}{\hat{r}}
+\frac{\left(\hat{q}_0^2 \hat{Q}^2\right) \left(M g^{t\phi}\right)^2}{\hat{r}^2}
+\hat{E}_0^2 \left(M g^{t\phi}\right)^2 \\
&-\frac{g^{tt} \left(\hat{q}_0^2 \hat{Q}^2\right) \left(M^2 g^{\phi\phi}\right)}{\hat{r}^2}
+\frac{2 \hat{E}_0 g^{tt} \left(\hat{q}_0 \hat{Q}\right) \left(M^2 g^{\phi\phi}\right)}{\hat{r}}
-\hat{E}_0^2 g^{tt} \left(M^2 g^{\phi\phi}\right).
\label{eq:pphi0}
\end{split}
\end{equation}

\begin{equation}\label{eq:E1}
\begin{split}
\hat{E}_1
&=\frac{\hat{q}_1 \hat{Q}}{\hat{r}}+ \frac{M g^{\text{t$\phi $}} \left(\hat{p}_{\text{$\phi $1}}-\frac{\hat{a} \hat{q}_1 \hat{Q}}{\hat{r}}\right)}{g^{\text{tt}}} \\
&-\frac{\sqrt{\left(M g^{\text{t$\phi $}}\right)^2 \left(\hat{p}_{\text{$\phi $1}}-\frac{\hat{a} \hat{q}_1 \hat{Q}}{\hat{r}}\right){}^2-g^{\text{tt}} \left(M^2 g^{\phi \phi } \left(\hat{p}_{\text{$\phi $1}}-\frac{\hat{a} \hat{q}_1 \hat{Q}}{\hat{r}}\right){}^2+1\right)}}{g^{\text{tt}}}
\end{split}
\end{equation}

The dimensionless mass ratio of the captured particle is obtained as
\begin{equation}
\tilde{\mu}_1
=
\frac{A+\sqrt{D_{\rm KN}}}{B},
\label{eq:mu1}
\end{equation}
where the coefficients $A$, $B$, and the discriminant $D_{\rm KN}$ are given by
\begin{subequations}
\begin{align}
A &= A\!\left(
\hat{E}_0,\hat{E}_1,
\hat{p}_{\phi0},\hat{p}_{\phi1},
\hat{q}_0,\hat{q}_1,
\hat{a},\hat{Q},\hat{r},
g^{tt},g^{t\phi},g^{\phi\phi}
\right), \\
B &= B\!\left(
\hat{E}_1,
\hat{p}_{\phi1},
\hat{q}_1,
\hat{a},\hat{Q},\hat{r},
g^{tt},g^{t\phi},g^{\phi\phi}
\right), \\
D_{\rm KN} &= D_{\rm KN}\!\left(
\hat{E}_0,\hat{E}_1,
\hat{p}_{\phi0},\hat{p}_{\phi1},
\hat{q}_0,\hat{q}_1,
\hat{a},\hat{Q},\hat{r},
g^{tt},g^{t\phi},g^{\phi\phi},\nu
\right).
\end{align}
\end{subequations}

Their explicit forms are presented in Appendix~\ref{app:coefficients}.

\begin{equation}\label{eq:E2q2}
\begin{split}
\hat{E}_2=\frac{\hat{E}_0-\hat{E}_1 \tilde{\mu }_1}{\nu  \tilde{\mu }_1}, \qquad  \hat{p}_{\text{$\phi $2}}=\frac{\hat{p}_{\text{$\phi $0}}-\tilde{\mu }_1 \hat{p}_{\text{$\phi $1}}}{\nu  \tilde{\mu }_1}  , \qquad \hat{q}_2=\frac{\hat{q}_0-\hat{q}_1 \tilde{\mu }_1}{\nu  \tilde{\mu }_1}.
\end{split}
\end{equation}

After each extraction event, the black hole mass, electric charge, and angular momentum are updated according to the conserved quantities carried away by the escaping and captured particles.

\begin{subequations}\label{eq:BH-update}
\begin{align}
M_n &= M_{n-1} + \mu_0\,\hat{E}_{1,n-1}\,\tilde{\mu}_{1,n-1}, \label{eq:Mn}\\
Q_n &= Q_{n-1} + \mu_0\,\hat{q}_{1,n-1}\,\tilde{\mu}_{1,n-1}, \label{eq:Qn}\\
L_n &= L_{n-1} + M_{n-1}\,\hat{p}_{\phi 1}\,\mu_{1,n-1}. \label{eq:Ln}
\end{align}
\end{subequations}
These changes induce a corresponding evolution of the dimensionless spin parameter
$\hat{a}_n = L_n/M_n^2$,
\begin{equation}\label{eq:delta-ahat}
\Delta\hat{a}_{n-1} = \frac{L_n}{M_n^2} - \frac{L_{n-1}}{M_{n-1}^2}.
\end{equation}
Accordingly, each extraction event modifies the outer horizon radius $(r_+)$, the irreducible mass $(M_{\rm irr})$, and the extractable energy $(E_{\rm extractable})$ through their dependence on the evolving black hole parameters. The total extracted energy is then obtained by cumulatively summing the energy extracted at each iteration.
\begin{equation}\label{extracted,n}
\begin{split}
E_{\mathrm{extracted},n}=M_0-M_n    
\end{split}
\end{equation}

To quantify the performance of the repetitive Penrose process, we employ two standard efficiency measures. The energy return on investment (EROI), defined as the ratio of the total extracted energy to the total energy of all incident particles supplied from infinity, is given by \cite{HuCaiWang2026}
\begin{equation}
\xi_n = \frac{E_{\rm extracted,n}}{n E_0},
\label{eq:EROI}
\end{equation}
where $E_{\rm extracted,n}$ is the cumulative energy extracted after $n$ iterations and $E_0$ is the energy of each incident particle.

The energy utilization efficiency (EUE), which measures the fraction of the \emph{reduced} extractable energy that is successfully transferred to escaping particles, is defined as \cite{HuCaiWang2026}
\begin{equation}
\Xi_n = \frac{E_{\rm extracted,n}}{E_{\rm extractable,0} - E_{\rm extractable,n}},
\label{eq:EUE}
\end{equation}
where $E_{\rm extractable,0}$ and $E_{\rm extractable,n}$ denote the extractable energy of the black hole before the first and after the $n$-th extraction event, respectively. The denominator $E_{\rm extractable,0} - E_{\rm extractable,n}$ is the total decrease in extractable energy, which is shared between the energy that escapes to infinity and the energy irreversibly deposited into $M_{\rm irr}$; hence $\Xi_n = 1$ corresponds to the reversible (Christodoulou) limit.

These quantities provide complementary insights: while the EROI can readily exceed unity (indicating net energy gain), the EUE is typically bounded well below 100\% due to the irreversible increase in the irreducible mass, consistent with the area theorem.

In a repetitive Penrose process, the sequence of energy-extraction events cannot continue indefinitely but is instead constrained by a set of physical conditions that determine the endpoint of the iteration. These conditions arise from both the kinematics of the particle decay and the geometry of the Kerr--Newman spacetime.

First, the decay must satisfy the mass-defect condition,
\begin{equation}
\mu_0-\mu_1-\mu_2>0,  \quad \rightarrow   \quad \tilde{ \mu}_1+  \tilde{ \mu}_2<1
\label{massdeficit}
\end{equation}
which corresponds to a positive mass deficit and ensures that the splitting process is kinematically admissible.

Second, the captured fragment must carry negative energy as measured by an observer at infinity,
\begin{equation}
\hat{E}_1<0,
\label{negativeenergy}
\end{equation}
since the extraction of black-hole energy relies on the absorption of negative energy through the event horizon. This requirement is the defining characteristic of the Penrose mechanism and can only be realized inside the ergoregion.

A further restriction arises from the structure of the effective potentials governing the particle trajectories. For particles $0$ and $2$, which must remain on trajectories connected to infinity, the classical turning points must be located on the outer (right) side of the maximum of the corresponding effective potential $V_i^{+}$. By contrast, the turning point of particle $1$, which is captured by the black hole, must lie on the inner (left) side of the potential barrier. The limiting configuration is reached when the turning point coincides exactly with the maximum of the effective potential. This critical condition is expressed as
\begin{equation}
\hat{V}_i^{+}(\hat r)=\hat E_i,
\qquad
\frac{d\hat V_i^{+}}{d\hat r}=0,
\label{criticalcondition}
\end{equation}
where $\hat r=r/M$ denotes the dimensionless decay radius. Physically, this double condition marks the onset of marginal stability for the corresponding orbit and therefore defines the boundary of the allowed parameter space.

For the particular case $\hat E_0=1$, corresponding to an incident particle released from rest at infinity, the lower bound on the black-hole spin is determined by the corotating marginally bound orbit of particle $0$. In the Kerr limit, one recovers the well-known result
\begin{equation}
\hat a_{\min,0}
=
2\sqrt{\hat r}
-
\hat r,
\label{aminKerr}
\end{equation}
which has been extensively discussed in the literature~\cite{Ruffini2025PRR}. In the Kerr--Newman spacetime, however, the presence of the electromagnetic interaction modifies the effective potential and consequently alters the minimum-spin condition. The resulting critical spin becomes a function of both the black-hole charge $\hat Q$ and the particle charge $\hat q_0$, leading to a richer parameter dependence than in the neutral Kerr case.

The repetitive Penrose process terminates when any of the above conditions can no longer be satisfied. As successive decay events extract energy and angular momentum from the black hole, the spacetime parameters evolve and the effective potentials are continuously deformed. Eventually, the critical condition (\ref{criticalcondition}) ceases to admit a physically acceptable solution, implying that no further decay capable of extracting energy can occur. This point defines the minimum spin threshold of the black hole and therefore marks the endpoint of the repetitive Penrose process, even though a finite amount of extractable energy may still remain stored in the spacetime.

\subsection{Minimum Spin Bound and Termination of the Repetitive Penrose Process}
To determine the endpoint of the repetitive Penrose process, we derive the minimum spin parameter for which the incident particle can still reach the splitting point and undergo a physically admissible decay. This critical value, denoted by $\hat{a}_{\min}$, defines the termination condition of the iterative energy-extraction sequence.

For particle~0, when $\hat{E}_0=1$, the minimum spin parameter of the black hole, $\hat{a}_{\min,0}$, is obtained by simultaneously solving Eqs.~\eqref{eq:pphi0}, \eqref{eq:E1}, \eqref{eq:mu1}, \eqref{eq:E2q2}, and \eqref{criticalcondition}. The resulting expression is
\begin{equation}\label{amin0}
\hat{a}_{\min,0} = \sqrt{\frac{A \sqrt{X} \sqrt{A^2 B^2 X-4 B Y Z+4 Y^2}}{2 \left(A^2 X-Z^2\right)}-\frac{A^2 B X}{2 \left(A^2 X-Z^2\right)}+\frac{Y Z}{A^2 X-Z^2}} .
\end{equation}
which represents the critical spin threshold below which particle~0 can no longer reach the decay radius,
where we have defined
\begin{subequations}\label{amin00}
\begin{align}
A &=\hat{q}_0 \hat{Q} \left(\hat{Q}^2-\hat{r}^2\right)+2 \hat{r} \left(\hat{r}-\hat{Q}^2\right) , \\
B &= \hat{Q}^2+\left(\hat{r}-2\right) \hat{r}, \\
X &=-2 \hat{q}_0 \hat{Q} \hat{r}+\left(\hat{q}_0^2-1\right) \hat{Q}^2+2 \hat{r} , \\
Y &=\left(\hat{Q}^2+\left(\hat{r}-2\right) \hat{r}\right) \left(-3 \hat{q}_0 \hat{Q}^3 \hat{r}-\left(\hat{q}_0^2-4\right) \hat{Q}^2 \hat{r}-\hat{q}_0 \hat{Q} \left(\hat{r}-4\right) \hat{r}^2+\left(\hat{q}_0^2-1\right) \hat{Q}^4+\left(\hat{r}-4\right) \hat{r}^2\right) , \\
Z &=\hat{q}_0^2 \hat{Q}^2 \left(\hat{Q}^2-\hat{r}^2\right)+\hat{q}_0 \hat{Q} \hat{r} \left(\hat{r} \left(\hat{r}+2\right)-3 \hat{Q}^2\right)-\left(\hat{Q}^2-\hat{r}\right) \left(\hat{Q}^2-\hat{r} \left(\hat{r}+2\right)\right) . 
\end{align}
\end{subequations}
In the neutral limit ($\hat{Q}=0$), this reduces to the known Kerr result $\hat{a}_{\rm min,0} = 2\sqrt{\hat{r}} - \hat{r}$. 
As is evident from Eqs.~\eqref{amin0} and~\eqref{amin00}, the critical spin parameter $\hat{a}_{\min,0}$ in the Kerr--Newman spacetime depends explicitly on both the black-hole charge $\hat{Q}$ and the charge of the incident particle $\hat{q}_0$. In contrast to the neutral Kerr case, where the minimum spin is determined solely by the spacetime geometry and the decay radius, the presence of electromagnetic interactions introduces an additional coupling between the black hole and the incoming particle. Consequently, the threshold spin required for particle~0 to reach the splitting radius is governed by the combined effects of gravitational attraction, frame dragging, and electrostatic interaction. Depending on the sign and magnitude of the product $\hat{Q}\hat{q}_0$, the electromagnetic force may either facilitate or hinder the particle's access to the ergoregion, thereby modifying the value of $\hat{a}_{\min,0}$.

In previous studies of the repetitive Penrose process, the threshold spin for terminating the iteration is typically determined by particle~2 whenever the incident particle has energy greater than its rest mass, i.e., $\hat{E}_0>1$. In this regime, the escaping particle becomes the limiting component of the process, and the corresponding minimum spin is associated with the corotating photon orbit.

In the Kerr--Newman spacetime, however, the presence of electromagnetic interactions between the black hole and the charged particles modifies this picture significantly. The electric force alters the effective potentials and consequently changes the relative importance of the constraints associated with particles~0, 1, and~2. Therefore, the threshold spin cannot be identified \emph{a priori} with the limiting condition of particle~2 alone.

To determine the termination point of the iterative process, we first calculate the minimum spin parameter associated with particle~2, denoted by $\hat{a}_{\min,2}$, which is obtained by imposing the marginal photon-orbit condition on the corotating photon sphere. Subsequently, this value is compared with the corresponding lower-spin limits of particles~0 and~1, namely $\hat{a}_{\min,0}$ and $\hat{a}_{\min,1}$. The physically relevant threshold spin is then identified as the most restrictive of these limits, since the repetitive Penrose process ceases as soon as any one of the participating particles can no longer satisfy its required orbital conditions.

Hence, unlike the neutral Kerr case, where particle~2 generally provides the dominant stopping criterion for $\hat{E}_0>1$, the Kerr--Newman case requires a simultaneous analysis of all three particles. The competition between gravitational and electromagnetic interactions may shift the limiting condition from one particle to another, thereby modifying the minimum spin at which the repetitive Penrose process terminates.

For particle~2, the limiting configuration occurs when its orbit approaches the corotating photon sphere of the Kerr--Newman black hole. The radius of the equatorial corotating photon orbit satisfies
\begin{equation}
\left(2\hat{Q}^{2}
\pm
2\hat{a}\sqrt{\hat{r}-\hat{Q}^{2}}\right)
+\hat{r}^{2}
-3\hat{r}
=0.
\end{equation}
Solving this relation for the spin parameter yields the corresponding lower spin limit
\begin{equation}
\hat{a}_{\min,2}
=
\frac{-2\hat{Q}^{2}-\hat{r}^{2}+3\hat{r}}
     {2\sqrt{\hat{r}-\hat{Q}^{2}}}.
\end{equation}

For particle~1, the limiting configuration is determined by the location of the event horizon. Since the captured particle must remain outside the horizon prior to crossing it, the threshold case is obtained when its turning point coincides with the outer horizon, i.e.
\begin{equation}
\Delta
=
\hat{a}^{2}
+\hat{Q}^{2}
+\hat{r}^{2}
-2\hat{r}
=
0.
\end{equation}
This immediately gives
\begin{equation}
\hat{a}_{\min,1}
=
\sqrt{
-\hat{Q}^{2}
-\hat{r}^{2}
+2\hat{r}
}.
\end{equation}

Finally, the actual spin threshold for the repetitive Penrose process is determined by comparing the three critical values
$\hat{a}_{\min,0}$,
$\hat{a}_{\min,1}$,
and
$\hat{a}_{\min,2}$.
The largest of these values defines the physical lower spin bound,
\begin{equation}
\hat{a}_{\min}
=
\max
\left\{
\hat{a}_{\min,0},
\hat{a}_{\min,1},
\hat{a}_{\min,2}
\right\},
\end{equation}
since the iterative extraction process can proceed only while all three particle trajectories remain physically admissible.

\begin{figure}[h!]
  \centering
  \begin{subfigure}[b]{0.48\textwidth}
    \centering
    \includegraphics[height=5cm,width=7cm]{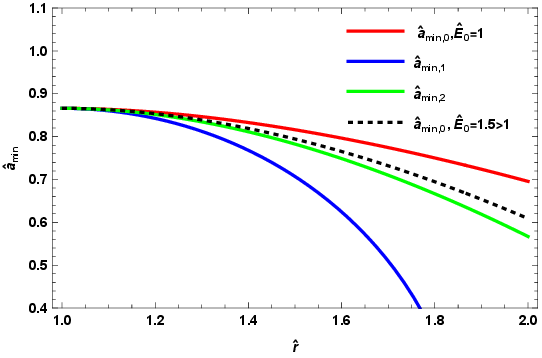}
    \caption{}
    \label{fig1amin}
  \end{subfigure}
  \hfill
  \begin{subfigure}[b]{0.48\textwidth}
    \centering
    \includegraphics[height=5cm,width=7cm]{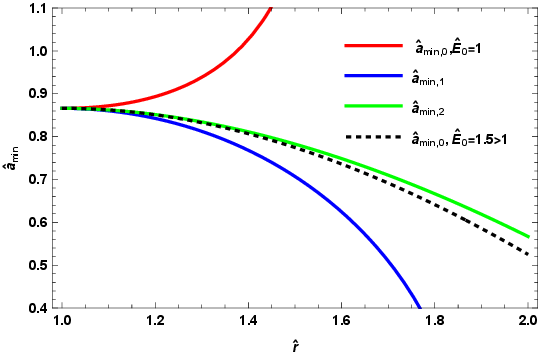}
    \caption{}
    \label{fig1amin1}
  \end{subfigure}
  \caption{\small{Minimum spin parameter $\hat{a}_{\min}$ as a function of the
  dimensionless decay radius $\hat{r}$, for $\hat{Q}=0.5$.
  \subref{fig1amin}~$\hat{q}_0=0.2$: gravity dominates
  ($\hat{Q}\hat{q}_0<1$) and $\hat{a}_{\min,0}$ governs termination.
  \subref{fig1amin1}~$\hat{q}_0=2.5$: electromagnetic repulsion dominates
  ($\hat{Q}\hat{q}_0\geq 1$) and $\hat{a}_{\min,2}$ governs termination.}}
  \label{m2}
\end{figure}

According to Figs.~\ref{fig1amin} and~\ref{fig1amin1}, and taking into account the electromagnetic interaction between the incoming particle (particle~0) and the black hole, the behavior of the critical spin limit strongly depends on the value of $\hat{Q}\hat{q}_0$. This quantity determines the relative strength of the electric repulsion compared to the gravitational attraction and plays a key role in establishing the dynamical regime of the process.
In the case $\hat{Q}\hat{q}_0 < 1$, the electric repulsion is weaker than the gravitational pull. Therefore, particle~0 can penetrate the ergosphere with its rest energy alone ($\hat{E}_0 = 1$) and decay at the splitting radius $\hat{r}$ into particles~1 and~2. In this regime, the main constraint on the iterative process comes from particle~0, and the stopping condition is governed by the minimum allowed spin for particle~0, denoted $\hat{a}_{\min,0}$ (shown in Fig.~\ref{fig1amin}). This lower bound marks the stability threshold for the orbit of the incoming particle; any further decrease in the spin would prevent the decay from occurring.
Conversely, when $\hat{Q}\hat{q}_0 \ge 1$, the electric repulsion overcomes the gravitational attraction and prevents particle~0 from reaching the decay region with rest energy alone. Under these circumstances, the incoming particle must be injected with a kinetic energy exceeding its rest mass ($\hat{E}_0 > 1$) in order to surmount the electric barrier and enter the ergosphere. In this regime, the relevant restriction shifts from particle~0 to the escaping particle~(particle~2), since the high energy of particle~2 and its escape conditions become the limiting factor for the continuation of the process. Consequently, the iterative stopping condition is now determined by the minimum spin allowed for particle~2, i.e. $\hat{a}_{\min,2}$ (depicted in Fig.~\ref{fig1amin1}).
This behavior highlights the crucial role of the ratio $\hat{Q}\hat{q}_0$ in determining the dynamical regime and in selecting the appropriate spin lower bound for terminating the repetitive decays.
In the following, we investigate the repetitive Penrose process. Reference~\cite{Ruffini2025PRR} indicates that choosing $\hat{E}_0 > 1$ results in a reduced EROI compared with the $\hat{E}_0 = 1$ case. Consequently, we select $\hat{E}_0 = 1$ in order to achieve the highest possible EROI.

\begin{figure}[h!]
  \centering
  \begin{subfigure}[b]{0.48\textwidth}
    \centering
    \includegraphics[height=5cm,width=7cm]{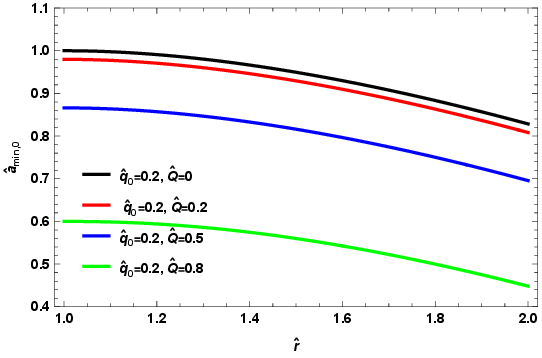}
    \caption{}
    \label{figamin0Q}
  \end{subfigure}
  \hfill
  \begin{subfigure}[b]{0.48\textwidth}
    \centering
    \includegraphics[height=5cm,width=7cm]{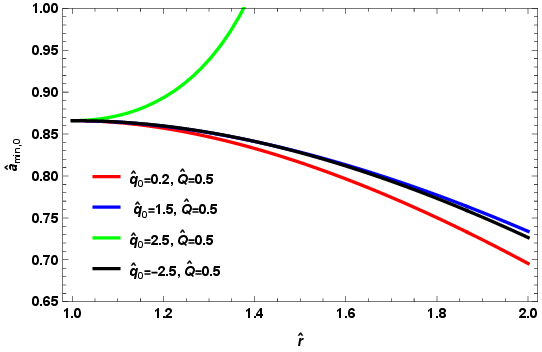}
    \caption{}
    \label{figamin0q0}
  \end{subfigure}
  \caption{\small{%
  Minimum spin parameter $\hat{a}_{\rm min,0}$ as a function of the
  dimensionless decay radius $\hat{r}$ for incident particles with $\hat{E}_0=1$.
  \subref{figamin0Q}~ Fixed $\hat{q}_0=0.2$ and different values of  $\hat{Q}$.
  \subref{figamin0q0}~ Fixed $\hat{Q}=0.5$ and different values of  $\hat{q}_0$.}}
  \label{amin0Qq0}
\end{figure}

Figure~\ref{amin0Qq0} shows the minimum spin parameter $\hat{a}_{\rm min,0}$ as a function of the dimensionless decay radius $\hat{r}$ for incident particles with $\hat{E}_0 = 1$.

In panel~\ref{figamin0Q}, for fixed $\hat{q}_0 = 0.2$, the minimum spin parameter decreases monotonically with increasing $\hat{r}$. Furthermore, increasing the black-hole charge $\hat{Q}$ systematically lowers the critical spin threshold. This behavior indicates that the electromagnetic interaction modifies the effective potential in a manner that reduces the critical spin required for the Penrose process to occur.

Panel~\ref{figamin0q0}  illustrates the dependence of $\hat{a}_{\rm min,0}$ on the incident particle charge $\hat{q}_0$ at fixed black-hole charge $\hat{Q}=0.5$. For moderate values of $\hat{q}_0$, the behavior remains qualitatively similar to that of the neutral case. However, for sufficiently large positive values of $\hat{q}_0$, $\hat{a}_{\rm min,0}$ increases rapidly, particularly at small decay radii. In this regime, the strong electromagnetic repulsion between the black hole and the incident particle raises the effective potential barrier, preventing particles released from rest at infinity ($\hat{E}_0=1$) from reaching the ergosphere with sufficient energy to produce a negative-energy fragment.

Overall, Fig.~\ref{amin0Qq0} demonstrates that the lower spin bound for the repetitive Penrose process is determined by the competition between gravitational attraction and electromagnetic interaction. The product $\hat{Q}\hat{q}_0$ therefore plays a crucial role in defining the allowed parameter space. For sufficiently small values of $\hat{Q}\hat{q}_0$, gravity dominates and the process can proceed efficiently. In contrast, when $\hat{Q}\hat{q}_0$ becomes large and positive, electromagnetic repulsion suppresses the process by increasing the minimum required spin.
 
 \newpage

\section{Numerical Implementation of the Repetitive Penrose Process in Kerr--Newman Black Holes} \label{sec:repetitive}
 
In this section, we present a detailed numerical analysis of the repetitive Penrose process for charged particles in the Kerr--Newman spacetime. To maximize the energy return on investment (EROI), we follow previous studies and set the incident energy to $\hat{E}_0 = 1$. This choice ensures that the incident particles arrive from rest at infinity, which generally leads to higher efficiency compared to cases with $\hat{E}_0 > 1$.

Following the methodology of Refs.~\cite{Ruffini2025PRR}, we adopt the following parameters for the incident particle and the decay: $\hat{p}_{\phi 1} = -19.434$, mass ratio $\nu = \mu_2 / \mu_1 = 0.78345$, and rest mass of the incident particle $\mu_0 = 10^{-2} M_0$. The decay is assumed to occur at the dimensionless radius $\hat{r} = 1.2$.

These values allow us to solve the analytic expressions for the decay products (Eqs.~(\ref{eq:pphi0})--(\ref{eq:E2q2})) at each iteration and update the black hole parameters $M_n$, $Q_n$, and $L_n$ accordingly. The results of this iterative procedure are presented in the following tables and figures. By examining different combinations of the black hole charge $\hat{Q}$, the incident particle charge $\hat{q}_0$, and the captured (infalling) particle charge $\hat{q}_1$, we systematically investigate how the electromagnetic coupling affects the energy extraction efficiency, the maximum number of iterations, and the final properties of the Kerr--Newman black hole.

 \begin{table}[htbp]
    \centering
    \setlength{\arrayrulewidth}{0.5pt}
    \renewcommand{\arraystretch}{1.2}
    \scriptsize
    \setlength{\tabcolsep}{3pt}
\resizebox{\textwidth}{!}{%
\begin{tabular}{ccccccccccccc}
\toprule
\multicolumn{13}{l}{\footnotesize $E_\text{extr}=E_{\text{extractable}}$;\; $E_\text{ext}=E_{\text{extracted}}$} \\
$n$ & $M/M_0$ & $\hat{a}_n$ & $\hat{Q}_n$ & $E_\text{extr}/M_0$ & $E_\text{ext}/M_0$ & $M_\text{irr}/M_0$ & $\tilde{\mu}_{1,n}$ & $\hat{E}_{1,n}$ & $\hat{p}_{\phi0,n}$ & $\hat{a}_{\min,0,n}$ & $\xi_n$ & $\Xi_n$  \\
\midrule
 0 & 1 & 1 & 0 & 0.292893 & 0 &  0.707107 & 0.0209006 & -6.93558 & 2.1127 & 0.99089 & 0 & 0 \\
 1 & 0.99855 & 0.998832 & 0 & 0.275611 & 0.00144958 & 0.72294 & 0.0208126 & -6.96513 & 2.12126 & 0.99089 & 0.144958 & 0.0838746 \\
 2 & 0.997101 & 0.997676 & 0 & 0.268419 & 0.0028992 & 0.728682 & 0.0207235 & -6.99529 & 2.13006 & 0.99089 & 0.14496 & 0.11846 \\
 3 & 0.995651 & 0.996532 & 0 & 0.262915 & 0.00434887 & 0.732736 & 0.0206333 & -7.0261 & 2.13914 & 0.99089 & 0.144962 & 0.145066 \\
 4 & 0.994201 & 0.995402 & 0 & 0.258294 & 0.00579858 & 0.735907 & 0.0205417 & -7.05764 & 2.14852 & 0.99089 & 0.144965 & 0.167593 \\
 5 & 0.992752 & 0.994284 & 0 & 0.254245 & 0.00724834 & 0.738506 & 0.0204488 & -7.08996 & 2.15822 & 0.99089 & 0.144967 & 0.187549 \\
 6 & 0.991302 & 0.99318 & 0 & 0.250608 & 0.00869815 & 0.740694 & 0.0203542 & -7.12314 & 2.16827 & 0.99089 & 0.144969 & 0.205702 \\
 7 & 0.989852 & 0.992089 & 0 & 0.247286 & 0.010148 & 0.742566 & 0.0202578 & -7.15727 & 2.17872 & 0.99089 & 0.144971 & 0.22251 \\
 8 & 0.988402 & 0.991013 & 0 & 0.244218 & 0.0115979 & 0.744184 & 0.0201593 & -7.19248 & 2.18961 & 0.99089 & 0.144974 & 0.238271 \\
 \textcolor{red}{9} & \textcolor{red}{0.986952} & \textcolor{red}{0.989952} & \textcolor{red}{0} & \textcolor{red}{0.24136} & \textcolor{red}{0.0130479} & \textcolor{red}{0.745592} & \textcolor{red}{0.0200585} & \textcolor{red}{-7.22889} & \textcolor{red}{2.20099} & \textcolor{red}{0.99089} & \textcolor{red}{0.144976} & \textcolor{red}{0.253193} \\
\bottomrule
\end{tabular}
}% end resizebox
\caption{Iterative evolution of the repetitive Penrose process for the neutral Kerr
baseline ($\hat{Q}_0=0$, $\hat{q}_0=\hat{q}_1=0$), which reproduces the known Kerr
results and serves as a reference. Fixed parameters: $\hat{E}_0=1$, $\hat{r}=1.2$,
$\nu=0.78345$, $\hat{p}_{\phi1}=-19.434$, $\mu_0=0.01M_0$, $M_0=1$.}
\label{P0}
\end{table}

\begin{table}[htbp]
    \centering
    \setlength{\arrayrulewidth}{0.5pt}
    \renewcommand{\arraystretch}{1.2}
    \scriptsize
    \setlength{\tabcolsep}{3pt}
\resizebox{\textwidth}{!}{%
\begin{tabular}{ccccccccccccc}
\toprule
$n$ & $M/M_0$ & $\hat{a}_n$ & $\hat{Q}_n$ & $E_\text{extr}/M_0$ & $E_\text{ext}/M_0$ & $M_\text{irr}/M_0$ & $\tilde{\mu}_{1,n}$ & $\hat{E}_{1,n}$ & $\hat{p}_{\phi0,n}$ & $\hat{a}_{\min,0,n}$ & $\xi_n$ & $\Xi_n$  \\
\midrule
0 & 1 & 0.866025 & 0.5 & 0.338562 & 0 & 0.661438 & 0.017247 & -6.412741 & 2.126825 & 0.85717& 0 & 0 \\
  1 & 0.998894 & 0.864585 & 0.5 & 0.319599 & 0.00110602 & 0.679295 & 0.0171494 & -6.44958 & 2.13816 & 0.85717& 0.110602 & 0.0583256 \\
 2 & 0.997788 & 0.863159 & 0.5 & 0.311776 & 0.00221209 & 0.686012 & 0.0170502 & -6.48734 & 2.14987 & 0.85717 & 0.110604 & 0.0825829\\
 3 & 0.996682 & 0.861748 & 0.5 & 0.305811 & 0.00331819 & 0.690871 & 0.0169495 & -6.52613 & 2.162 & 0.85717& 0.110606 & 0.101316 \\
 4 & 0.995576 & 0.860351 & 0.5 & 0.300819 & 0.00442434 & 0.694757 & 0.0168471 & -6.56602 & 2.17459 & 0.85717& 0.110608 & 0.117222 \\
 5 & 0.994469 & 0.85897 & 0.5 & 0.296455 & 0.00553052 & 0.698014 & 0.0167429 & -6.60715 & 2.18768 & 0.85717& 0.11061 & 0.131344 \\
 6 & 0.993363 & 0.857605 & 0.5 & 0.292542 & 0.00663675 & 0.700821 & 0.0166365 & -6.64964 & 2.20133 & 0.85717& 0.110613 & 0.144215 \\
\textcolor{red}{ 7} & \textcolor{red}{0.992257} & \textcolor{red}{0.856257} & \textcolor{red}{0.5} & \textcolor{red}{0.288975} & \textcolor{red}{0.00774302} & \textcolor{red}{0.703282} & \textcolor{red}{0.0165277} & \textcolor{red}{-6.69366} & \textcolor{red}{2.21561} & \textcolor{red}{0.85717}& \textcolor{red}{0.110615} & \textcolor{red}{0.156151} \\
\bottomrule
\end{tabular}
}% end resizebox
\caption{Repetitive Penrose process for neutral particles ($\hat{q}_0=\hat{q}_1=0$) in a
charged Kerr--Newman black hole with $\hat{Q}_0=0.5$, isolating the effect of the horizon
charge alone. Since no charge is transferred, $\hat{Q}_n=0.5$ remains constant throughout.
Fixed parameters as in Table~\ref{P0}.}
\label{P1}
\end{table}

\begin{table}[htbp]
    \centering
    \setlength{\arrayrulewidth}{0.5pt}
    \renewcommand{\arraystretch}{1.2}
    \scriptsize
    \setlength{\tabcolsep}{3pt}
\resizebox{\textwidth}{!}{%
\begin{tabular}{ccccccccccccc}
\toprule
$n$ & $M/M_0$ & $\hat{a}_n$ & $\hat{Q}_n$ & $E_\text{extr}/M_0$ & $E_\text{ext}/M_0$ & $M_\text{irr}/M_0$ & $\tilde{\mu}_{1,n}$ & $\hat{E}_{1,n}$ & $\hat{p}_{\phi0,n}$ & $\hat{a}_{\min,0,n}$ & $\xi_n$ & $\Xi_n$ \\
\midrule
 0 & 1 & 0.866025 & 0.5 & 0.338562 & 0 & 0.661438 & 0.016935 & -4.925666 & 2.007715 & 0.857057 & 0 & 0\\
  1 & 0.999166 & 0.864175 & 0.500847 & 0.320362 & 0.000834172 & 0.678804 & 0.0168352 & -4.95728 & 2.01769 & 0.856568& 0.0834172 & 0.045833 \\
2 & 0.998331 & 0.862341 & 0.501689 & 0.312985 & 0.00166874 & 0.685346 & 0.0167343 & -4.98979 & 2.02798 & 0.856082 & 0.0834371 & 0.0652432 \\
 3 & 0.997496 & 0.860522 & 0.502525 & 0.307412 & 0.00250375 & 0.690085 & 0.0166321 & -5.02327 & 2.03861 & 0.855598 & 0.0834582 & 0.0803756 \\
 4 & 0.996661 & 0.85872 & 0.503357 & 0.302781 & 0.00333922 & 0.69388 & 0.0165288 & -5.05779 & 2.04962 & 0.855115 & 0.0834806 & 0.0933232 \\
 5 & 0.995825 & 0.856934 & 0.504183 & 0.298758 & 0.00417521 & 0.697067 & 0.016424 & -5.09347 & 2.06103 & 0.854634 & 0.0835043 & 0.104895 \\
 6 & 0.994988 & 0.855165 & 0.505004 & 0.295172 & 0.00501176 & 0.699816 & 0.0163176 & -5.1304 & 2.07289 & 0.854156 & 0.0835294 & 0.115504 \\
 \textcolor{red}{7} & \textcolor{red}{0.994151} & \textcolor{red}{0.853413} & \textcolor{red}{0.50582} & \textcolor{red}{0.291919} & \textcolor{red}{0.00584892} & \textcolor{red}{0.702232} & \textcolor{red}{0.0162094} & \textcolor{red}{-5.16872} & \textcolor{red}{2.08524} & \textcolor{red}{0.853679} & \textcolor{red}{0.083556} & \textcolor{red}{0.125397}\\
\bottomrule
\end{tabular}
}% end resizebox
\caption{Repetitive Penrose process with $\hat{q}_0=0.2$ and $\hat{q}_1=+5$
(captured fragment carries the same sign of charge as the horizon;
$\hat{Q}\hat{q}_1=2.5>0$, repulsive electromagnetic coupling),
$\hat{Q}_0=0.5$. This configuration weakens the negative-energy channel relative
to the attractively coupled case ($\hat{q}_1<0$).
Fixed parameters as in Table~\ref{P0}.}
\label{P2}
\end{table}

\begin{table}[htbp]
    \centering
    \setlength{\arrayrulewidth}{0.5pt}
    \renewcommand{\arraystretch}{1.2}
    \scriptsize
    \setlength{\tabcolsep}{3pt}
\resizebox{\textwidth}{!}{%
\begin{tabular}{ccccccccccccc}
\toprule
$n$ & $M/M_0$ & $\hat{a}_n$ & $\hat{Q}_n$ & $E_\text{extr}/M_0$ & $E_\text{ext}/M_0$ & $M_\text{irr}/M_0$ & $\tilde{\mu}_{1,n}$ & $\hat{E}_{1,n}$ & $\hat{p}_{\phi0,n}$ & $\hat{a}_{\min,0,n}$ & $\xi_n$ & $\Xi_n$ \\
\midrule
 0 & 1 & 0.866025 & 0.5 & 0.338562 & 0 & 0.661438 & 0.017783 & -7.899726 & 2.245171 & 0.857321& 0 & 0 \\
 1 & 0.998595 & 0.864998 & 0.499111 & 0.318712 & 0.00140484 & 0.679883 & 0.0176857 & -7.93582 & 2.2576 & 0.857831 & 0.140484 & 0.0707723 \\
 2 & 0.997192 & 0.863983 & 0.498227 & 0.310391 & 0.00280834 & 0.6868 & 0.0175863 & -7.9729 & 2.27048 & 0.858337 & 0.140417 & 0.0996895\\
 3 & 0.99579 & 0.862981 & 0.497347 & 0.303999 & 0.00421047 & 0.69179 & 0.0174849 & -8.01107 & 2.28387 & 0.858838 & 0.140349 & 0.12182 \\
 4 & 0.994389 & 0.861992 & 0.496473 & 0.298618 & 0.0056112 & 0.695771 & 0.0173815 & -8.05043 & 2.29782 & 0.859336 & 0.14028 & 0.140476\\
 5 & 0.99299 & 0.861016 & 0.495604 & 0.293891 & 0.00701049 & 0.699098 & 0.0172756 & -8.09114 & 2.31239 & 0.859829 & 0.14021 & 0.156937 \\
\textcolor{red}{ 6} & \textcolor{red}{0.991592} & \textcolor{red}{0.860055} & \textcolor{red}{0.49474} & \textcolor{red}{0.289636} & \textcolor{red}{0.00840828} & \textcolor{red}{0.701955} & \textcolor{red}{0.0171668} & \textcolor{red}{-8.13335} & \textcolor{red}{2.32767} & \textcolor{red}{0.860319} & \textcolor{red}{0.140138} & \textcolor{red}{0.171858}\\
\bottomrule
\end{tabular}
}% end resizebox
\caption{Repetitive Penrose process with $\hat{q}_0=-0.2$ and $\hat{q}_1=-5$
(oppositely charged incident particle, attractive coupling $\hat{Q}\hat{q}_1<0$),
$\hat{Q}_0=0.5$. Fixed parameters as in Table~\ref{P0}.}
\label{P3}
\end{table}

\begin{table}[htbp]
    \centering
    \setlength{\arrayrulewidth}{0.5pt}
    \renewcommand{\arraystretch}{1.2}
    \scriptsize
    \setlength{\tabcolsep}{3pt}
\resizebox{\textwidth}{!}{%
\begin{tabular}{ccccccccccccc}
\toprule
$n$ & $M/M_0$ & $\hat{a}_n$ & $\hat{Q}_n$ & $E_\text{extr}/M_0$ & $E_\text{ext}/M_0$ & $M_\text{irr}/M_0$ & $\tilde{\mu}_{1,n}$ & $\hat{E}_{1,n}$ & $\hat{p}_{\phi0,n}$ & $\hat{a}_{\min,0,n}$ & $\xi_n$ & $\Xi_n$\\
\midrule
0 & 1 & 0.866025 & 0.5 & 0.338562 & 0 & 0.661438 & 0.020383 & -7.899726 & 2.007715 & 0.857057&  0 & 0 \\
1 & 0.99839 & 0.864847 & 0.498981 & 0.317284 & 0.00161021 & 0.681106 & 0.0202482 & -7.94123 & 2.02107 & 0.857643 & 0.161021 & 0.0756747\\
2 & 0.996782 & 0.863685 & 0.497968 & 0.308369 & 0.00321816 & 0.688413 & 0.0201108 & -7.98403 & 2.03498 & 0.858223& 0.160908 & 0.106585 \\
3 & 0.995176 & 0.862541 & 0.496963 & 0.301527 & 0.00482381 & 0.693649 & 0.0199706 & -8.0283 & 2.04949 & 0.858798 & 0.160794 & 0.13025\\
4 & 0.993573 & 0.861415 & 0.495964 & 0.295775 & 0.00642711 & 0.697797 & 0.0198271 & -8.07421 & 2.0647 & 0.859368 & 0.160678 & 0.150213\\
5 & 0.991972 & 0.860307 & 0.494973 & 0.290731 & 0.00802799 & 0.701241 & 0.0196798 & -8.12199 & 2.0807 & 0.859932 & 0.16056 & 0.167839\\
\textcolor{red}{6} & \textcolor{red}{0.990374} & \textcolor{red}{0.859218} & \textcolor{red}{0.493989} & \textcolor{red}{0.286197} & \textcolor{red}{0.00962637} & \textcolor{red}{0.704177} & \textcolor{red}{0.0195279} & \textcolor{red}{-8.17194} & \textcolor{red}{2.09761} & \textcolor{red}{0.86049} & \textcolor{red}{0.16044} & \textcolor{red}{0.183831}  \\
\bottomrule
\end{tabular}
}% end resizebox
\caption{Repetitive Penrose process for the optimal charge configuration
($\hat{Q}\hat{q}_0>0$, $\hat{Q}\hat{q}_1<0$) with $\hat{q}_0=0.2$ and $\hat{q}_1=-5$,
$\hat{Q}_0=0.5$. Fixed parameters as in Table~\ref{P0}.}
\label{P4}
\end{table}

\begin{table}[htbp]
    \centering
    \setlength{\arrayrulewidth}{0.5pt}
    \renewcommand{\arraystretch}{1.2}
    \scriptsize
    \setlength{\tabcolsep}{3pt}
\resizebox{\textwidth}{!}{%
\begin{tabular}{ccccccccccccc}
\toprule
$n$ & $M/M_0$ & $\hat{a}_n$ & $\hat{Q}_n$ & $E_\text{extr}/M_0$ & $E_\text{ext}/M_0$ & $M_\text{irr}/M_0$ & $\tilde{\mu}_{1,n}$ & $\hat{E}_{1,n}$ & $\hat{p}_{\phi0,n}$ & $\hat{a}_{\min,0,n}$ & $\xi_n$ & $\Xi_n$\\
\midrule
 0 & 1 & 0.866025 & 0.5 & 0.338562 & 0 & 0.661438 & 0.0263034 & -11.1914 & 2.007715 & 0.857057&  0 & 0 \\
 1 & 0.997056 & 0.866005 & 0.495773 & 0.312688 & 0.00294373 & 0.684368 & 0.0260305 & -11.204 & 2.02686 & 0.859477 & 0.294373 & 0.113772 \\
 2 & 0.99414 & 0.865990 & 0.49159 & 0.301506 & 0.00586019 & 0.692634 & 0.0257535 & -11.2189 & 2.04698 & 0.861845 & 0.29301 & 0.158145 \\
 3 & 0.991251 & 0.865982 & 0.487451 & 0.29284 & 0.00874946 & 0.69841 & 0.025471 & -11.2365 & 2.06826 & 0.864163 & 0.291649 & 0.191362 \\
 \textcolor{red}{4} & \textcolor{red}{0.988388} & \textcolor{red}{0.865981} & \textcolor{red}{0.483358} & \textcolor{red}{0.285518} & \textcolor{red}{0.0116115} & \textcolor{red}{0.702871} & \textcolor{red}{0.0251812} & \textcolor{red}{-11.2573} & \textcolor{red}{2.09094} & \textcolor{red}{0.86643} & \textcolor{red}{0.290288} & \textcolor{red}{0.218902} \\
\bottomrule
\end{tabular}
}% end resizebox
\caption{Repetitive Penrose process at $\hat{q}_1=-16.07$ (Region~I), marking the
threshold beyond which the dimensionless spin $\hat{a}_n$ begins to grow despite the
decreasing total angular momentum; $\hat{q}_0=0.2$, $\hat{Q}_0=0.5$. Fixed parameters as
in Table~\ref{P0}.}
\label{P-16.07}
\end{table}

\begin{table}[htbp]
    \centering
    \setlength{\arrayrulewidth}{0.5pt}
    \renewcommand{\arraystretch}{1.2}
    \scriptsize
    \setlength{\tabcolsep}{3pt}
\resizebox{\textwidth}{!}{%
\begin{tabular}{ccccccccccccc}
\toprule
$n$ & $M/M_0$ & $\hat{a}_n$ & $\hat{Q}_n$ & $E_\text{extr}/M_0$ & $E_\text{ext}/M_0$ & $M_\text{irr}/M_0$ & $\tilde{\mu}_{1,n}$ & $\hat{E}_{1,n}$ & $\hat{p}_{\phi0,n}$ & $\hat{a}_{\min,0,n}$ & $\xi_n$ & $\Xi_n$\\
\midrule
 0 & 1 & 0.866025 & 0.5 & 0.338562 & 0 & 0.661438 & 0.094992 & -19.4844 & 2.007715 & 0.857057&  0 & 0 \\
 1 & 0.981491 & 0.879833 & 0.458192 & 0.280106 & 0.0185094 & 0.701384 & 0.0735862 & -18.4099 & 2.09102 & 0.879831 & 1.85094 & 0.316638 \\
 \textcolor{red}{2} & \textcolor{red}{0.967943} & \textcolor{red}{0.889652} & \textcolor{red}{0.425807} & \textcolor{red}{0.258535} & \textcolor{red}{0.0320566} & \textcolor{red}{0.709408} & \textcolor{red}{0.0615138} & \textcolor{red}{-17.6258} & \textcolor{red}{2.17028} & \textcolor{red}{0.895763} & \textcolor{red}{1.60283} & \textcolor{red}{0.400573} \\
\bottomrule
\end{tabular}
}% end resizebox
\caption{Repetitive Penrose process at $\hat{q}_1=-44.01$ (upper edge of Region~I),
the most negative captured charge for which the iteration is still sustained;
$\hat{q}_0=0.2$, $\hat{Q}_0=0.5$. Fixed parameters as in Table~\ref{P0}.}
\label{P-44.01}
\end{table}

\begin{table}[htbp]
    \centering
    \setlength{\arrayrulewidth}{0.5pt}
    \renewcommand{\arraystretch}{1.2}
    \scriptsize
    \setlength{\tabcolsep}{3pt}
\resizebox{\textwidth}{!}{%
\begin{tabular}{ccccccccccccc}
\toprule
$n$ & $M/M_0$ & $\hat{a}_n$ & $\hat{Q}_n$ & $E_\text{extr}/M_0$ & $E_\text{ext}/M_0$ & $M_\text{irr}/M_0$ & $\tilde{\mu}_{1,n}$ & $\hat{E}_{1,n}$ & $\hat{p}_{\phi0,n}$ & $\hat{a}_{\min,0,n}$ & $\xi_n$ & $\Xi_n$\\
\midrule
 0 & 1 & 0.866025 & 0.5 & 0.338562 & 0 & 0.661438 & 0.380574 & -22.4303 & 2.007715 & 0.857057&  0 & 0 \\
 1 & 0.914636 & 0.946812 & 0.293653 & 0.239889 & 0.0853638 & 0.674748 & 0.0546222 & -15.7513 & 2.12567 & 0.946811 & 8.53638 & 0.865113 \\
 \textcolor{red}{2} & \textcolor{red}{0.906033} & \textcolor{red}{0.953052} & \textcolor{red}{0.264037} & \textcolor{red}{0.230025} & \textcolor{red}{0.0939674} & \textcolor{red}{0.676008} & \textcolor{red}{0.0466456} & \textcolor{red}{-14.8956} & \textcolor{red}{2.15795} & \textcolor{red}{0.955408} & \textcolor{red}{4.69837} & \textcolor{red}{0.86576} \\
\bottomrule
\end{tabular}
}% end resizebox
\caption{Repetitive Penrose process at $\hat{q}_1=-54.22$ (lower edge of Region~III),
where the process reappears and both EROI and EUE rise sharply; $\hat{q}_0=0.2$,
$\hat{Q}_0=0.5$. Fixed parameters as in Table~\ref{P0}.}
\label{P-54.22}
\end{table}

\begin{table}[htbp]
    \centering
    \setlength{\arrayrulewidth}{0.5pt}
    \renewcommand{\arraystretch}{1.2}
    \scriptsize
    \setlength{\tabcolsep}{3pt}
\resizebox{\textwidth}{!}{%
\begin{tabular}{ccccccccccccc}
\toprule
$n$ & $M/M_0$ & $\hat{a}_n$ & $\hat{Q}_n$ & $E_\text{extr}/M_0$ & $E_\text{ext}/M_0$ & $M_\text{irr}/M_0$ & $\tilde{\mu}_{1,n}$ & $\hat{E}_{1,n}$ & $\hat{p}_{\phi0,n}$ & $\hat{a}_{\min,0,n}$ & $\xi_n$ & $\Xi_n$ \\
\midrule
 0 & 1 & 0.866025 & 0.5 & 0.338562 & 0 & 0.661438 & 0.417586 & -22.603250 & 2.007715 & 0.857057 & 0 & 0\\
 1 & 0.905612 & 0.957005 & 0.270937 & 0.244174 & 0.0943881 & 0.661438 & 0.0507822 & -15.092 & 2.09778 & 0.953493 & 9.43881 & 0.999999 \\
 2 & 0.897948 & 0.962327 & 0.243081 & 0.234356 & 0.102052 & 0.663592 & 0.0441337 & -14.2766 & 2.1229 & 0.9609 & 5.10261 & 0.97933 \\
 \textcolor{red}{3} & \textcolor{red}{0.891647} & \textcolor{red}{0.966288} & \textcolor{red}{0.218872} & \textcolor{red}{0.226898} & \textcolor{red}{0.108353} & \textcolor{red}{0.664749} & \textcolor{red}{0.0394736} & \textcolor{red}{-13.5829} & \textcolor{red}{2.1461} & \textcolor{red}{0.966646} & \textcolor{red}{3.61176} & \textcolor{red}{0.970346} \\

\bottomrule
\end{tabular}
}% end resizebox
\caption{Repetitive Penrose process at the critical charge $\hat{q}_1=-54.85405$ (upper
boundary of Region~III), where the irreducible mass is nearly stationary
($\Delta M_{\mathrm{irr}}\approx0$) and the process operates at the reversible thermodynamic limit; $\hat{q}_0=0.2$, $\hat{Q}_0=0.5$. Fixed parameters as in
Table~\ref{P0}.}
\label{P-54.85405}
\end{table}

\begin{table}[htbp]
    \centering
    \setlength{\arrayrulewidth}{0.5pt}
    \renewcommand{\arraystretch}{1.2}
    \scriptsize
    \setlength{\tabcolsep}{3pt}
\resizebox{\textwidth}{!}{%
\begin{tabular}{ccccccccccccc}
\toprule
$n$ & $M/M_0$ & $\hat{a}_n$ & $\hat{Q}_n$ & $E_\text{extr}/M_0$ & $E_\text{ext}/M_0$ & $M_\text{irr}/M_0$ & $\tilde{\mu}_{1,n}$ & $\hat{E}_{1,n}$ & $\hat{p}_{\phi0,n}$ & $\hat{a}_{\min,0,n}$ & $\xi_n$ & $\Xi_n$\\
\midrule
 0 & 1 & 0.866025 & 0.5 & 0.338562 & 0 & 0.661438 & 0.418055 & -22.6049 & 2.007715 & 0.857057&  0 & 0 \\
 1 & 0.905527 & 0.957103 & 0.270723 & 0.244238 & 0.0944727 & 0.661289 & 0.0507485 & -15.0857 & 2.09751 & 0.953553 & 9.44727 & 1.00158 \\
 2 & 0.897872 & 0.962416 & 0.242883 & 0.234415 & 0.102128 & 0.663457 & 0.044111 & -14.2706 & 2.12257 & 0.96095 & 5.10642 & 0.980616 \\
 \textcolor{red}{3} & \textcolor{red}{0.891577} & \textcolor{red}{0.966371} & \textcolor{red}{0.218683} & \textcolor{red}{0.226953} & \textcolor{red}{0.108423} & \textcolor{red}{0.664623} & \textcolor{red}{0.0394571} & \textcolor{red}{-13.5771} & \textcolor{red}{2.14572} & \textcolor{red}{0.966688} & \textcolor{red}{3.61411} & \textcolor{red}{0.971458} \\
\bottomrule
\end{tabular}
}% end resizebox
\caption{Repetitive Penrose process at $\hat{q}_1=-54.86$ (Region~IV): the kinematic
conditions are met, but $\Delta M_{\mathrm{irr}}<0$ violates the Hawking area theorem,
signaling the breakdown of the test-particle approximation; $\hat{q}_0=0.2$,
$\hat{Q}_0=0.5$. Fixed parameters as in Table~\ref{P0}.}
\label{P-54.86}
\end{table}
\newpage

\begin{figure}[h!]
  \centering
  \begin{subfigure}[b]{0.32\textwidth}
    \centering
    \includegraphics[height=4.5cm,width=4.5cm]{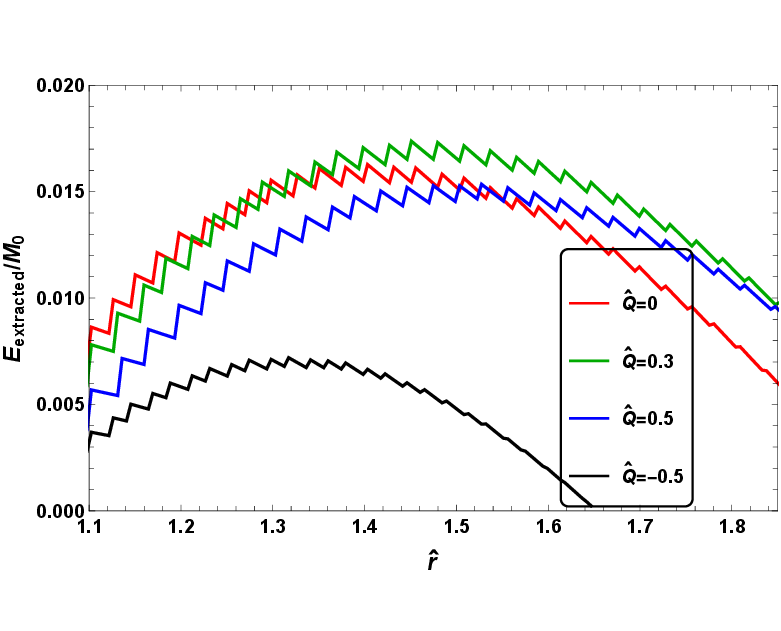}
    \caption{}
    \label{figextractedQ}
  \end{subfigure}
  \hfill
  \begin{subfigure}[b]{0.32\textwidth}
    \centering
    \includegraphics[height=4.5cm,width=4.5cm]{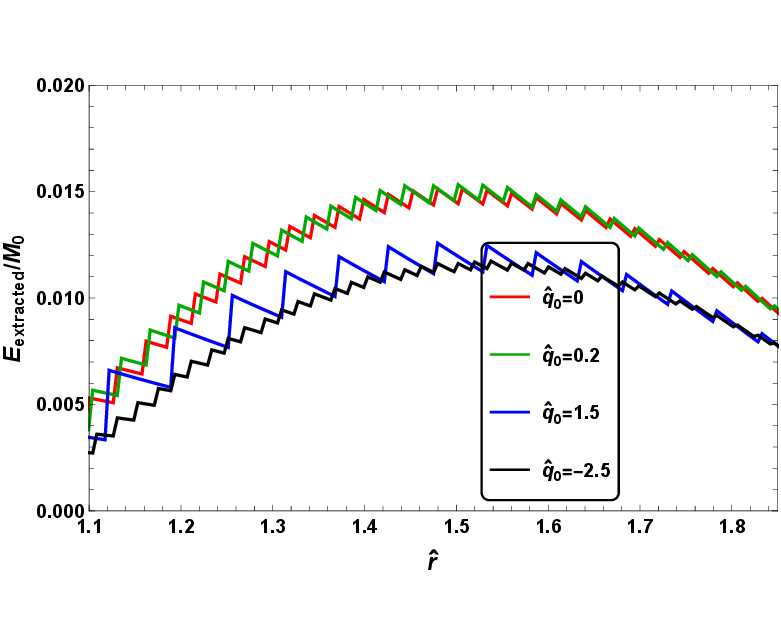}
    \caption{}
    \label{figextractedq0}
  \end{subfigure}
  \hfill
  \begin{subfigure}[b]{0.32\textwidth}
    \centering
    \includegraphics[height=4.5cm,width=4.5cm]{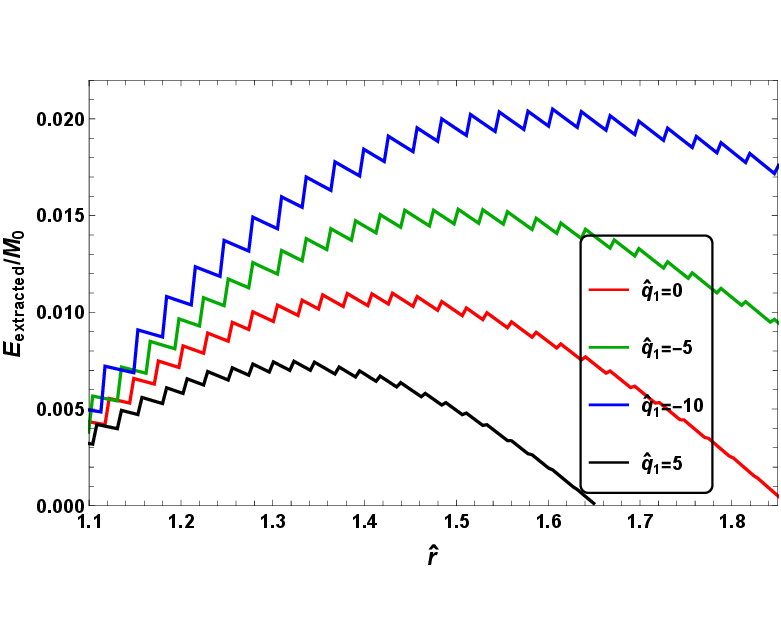}
    \caption{}
    \label{figextractedq1}
  \end{subfigure}
 \caption{
Final extracted energy $E_{\rm extracted}/M_0$ of the repetitive Penrose process
as a function of the dimensionless decay radius $\hat r$.
Each point corresponds to the final iteration $n_f(\hat r)$ at which the process
terminates.
Panel~\subref{figextractedQ} varies the black-hole charge $\hat Q$,
panel~\subref{figextractedq0} the incident-particle charge $\hat q_0$,
and panel~\subref{figextractedq1} the captured-particle charge $\hat q_1$.
The remaining parameters are
$\hat E_0=1$,
$\nu=0.78345$,
$\hat p_{\phi1}=-19.434$,
$\mu_0=10^{-2}M_0$,
and $M_0=1$.
The staircase-like modulation originates from the discrete variation of the terminal
iteration number $n_f$ with $\hat r$.
}
  \label{fig:Eext}
\end{figure}

\begin{figure}[h!]
  \centering
  \begin{subfigure}[b]{0.32\textwidth}
    \centering
    \includegraphics[height=4.5cm,width=4.5cm]{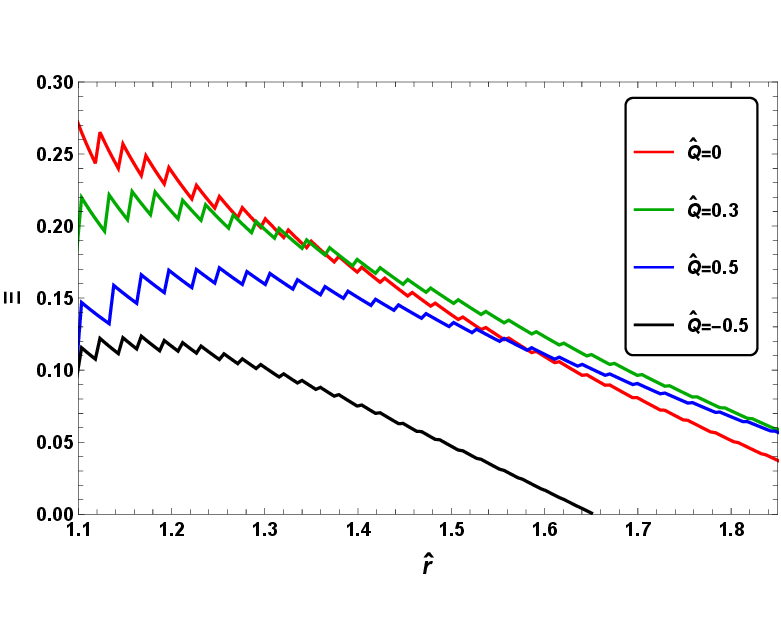}
    \caption{}
    \label{figexiQ}
  \end{subfigure}
  \hfill
  \begin{subfigure}[b]{0.32\textwidth}
    \centering
    \includegraphics[height=4.5cm,width=4.5cm]{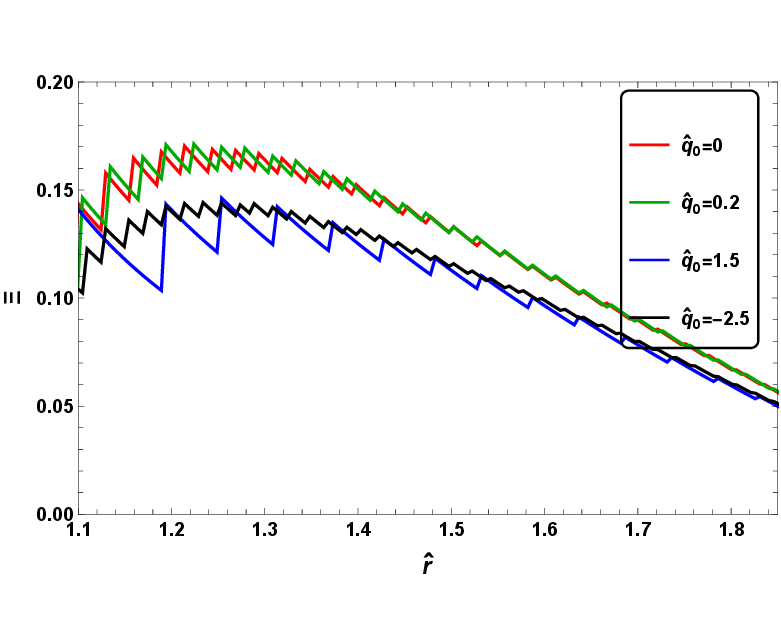}
    \caption{}
    \label{figexiq0}
  \end{subfigure}
  \hfill
  \begin{subfigure}[b]{0.32\textwidth}
    \centering
    \includegraphics[height=4.5cm,width=4.5cm]{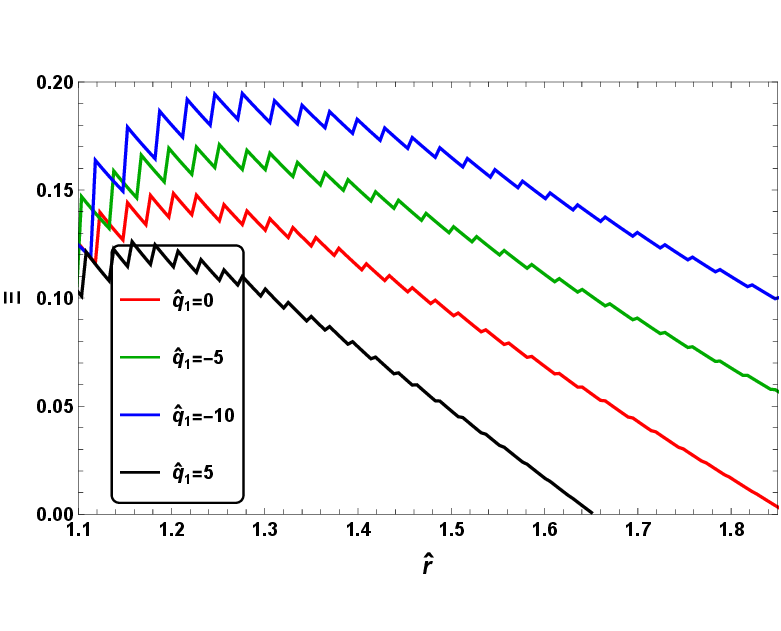}
    \caption{}
    \label{figexiq1}
  \end{subfigure}
  \caption{
Final energy utilization efficiency (EUE), $\Xi$, as a function of the decay radius
$\hat r$ for the same parameter sets as Fig.~\ref{fig:Eext}.
Panels~\subref{figexiQ}--\subref{figexiq1} correspond to variations of the
black-hole charge $\hat Q$, the incident-particle charge $\hat q_0$, and the
captured-particle charge $\hat q_1$, respectively.
}
  \label{fig:EUE}
\end{figure}

\begin{figure}[h!]
  \centering
  \begin{subfigure}[b]{0.32\textwidth}
    \centering
    \includegraphics[height=4.5cm,width=4.5cm]{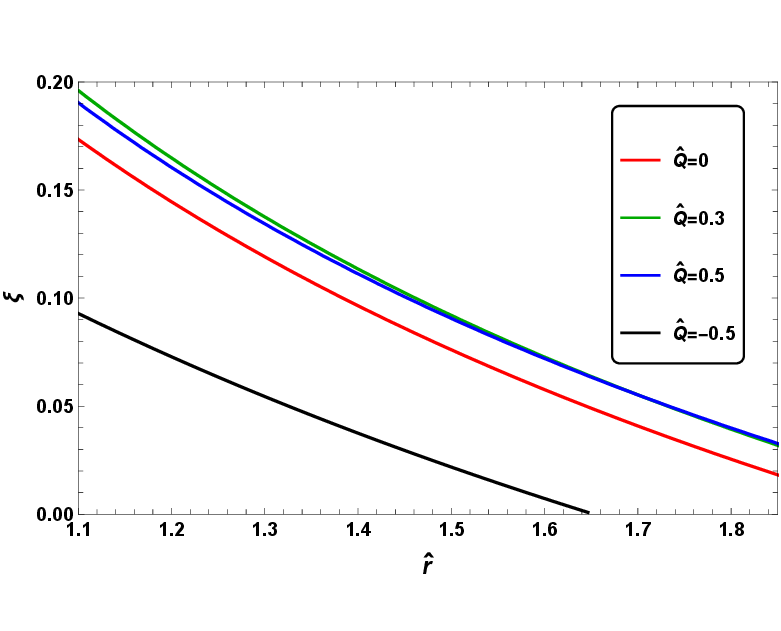}
    \caption{}
    \label{figeroiQ}
  \end{subfigure}
  \hfill
  \begin{subfigure}[b]{0.32\textwidth}
    \centering
    \includegraphics[height=4.5cm,width=4.5cm]{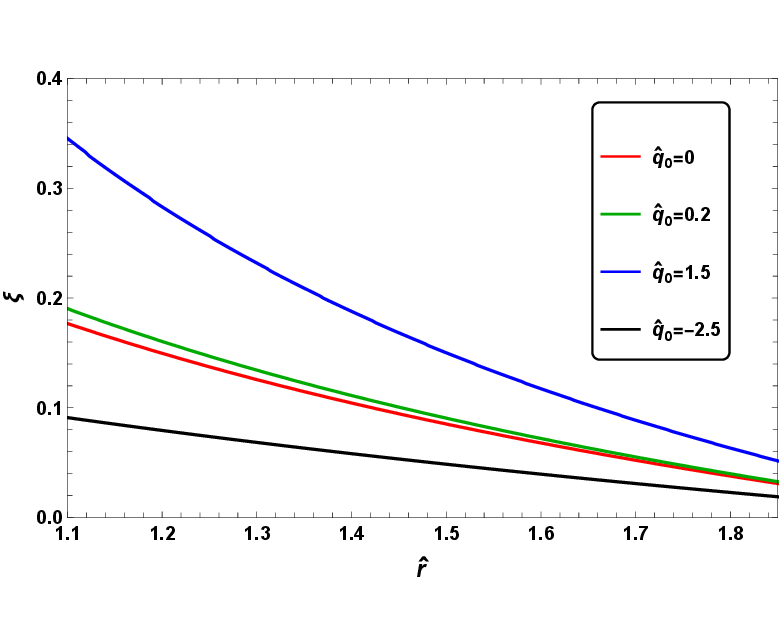}
    \caption{}
    \label{figeroiq0}
  \end{subfigure}
  \hfill
  \begin{subfigure}[b]{0.32\textwidth}
    \centering
    \includegraphics[height=4.5cm,width=4.5cm]{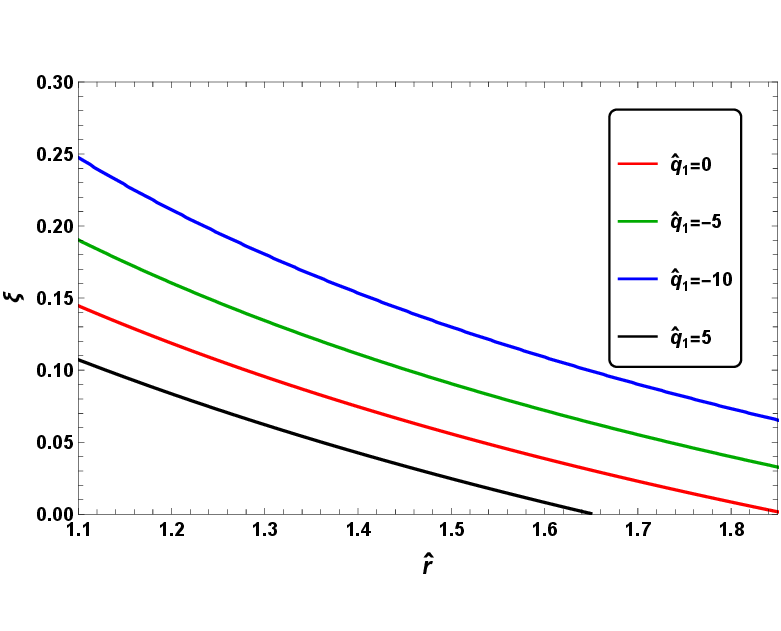}
    \caption{}
    \label{figeroiq1}
  \end{subfigure}
\caption{
Final energy return on investment (EROI), $\xi$, as a function of the decay radius
$\hat r$ for the same parameter sets as Fig.~\ref{fig:Eext}.
Panels~\subref{figeroiQ}--\subref{figeroiq1} correspond to variations of the
black-hole charge $\hat Q$, the incident-particle charge $\hat q_0$, and the
captured-particle charge $\hat q_1$, respectively.
}
  \label{fig:EROI}
\end{figure}

Figures~\ref{fig:Eext}, \ref{fig:EUE}, and \ref{fig:EROI} show the total extracted
energy $E_{\rm extracted}/M_0$, the energy utilization efficiency (EUE) $\Xi$, and
the energy return on investment (EROI) $\xi$ as functions of the dimensionless decay
radius $\hat r$, respectively. Each curve represents the final state of the repetitive
Penrose process, with the plotted value evaluated at the terminal iteration
$n_f(\hat r)$ corresponding to that decay radius. Panel~(a) varies the black-hole
charge $\hat Q$, panel~(b) the incident-particle charge $\hat q_0$, and
panel~(c) the captured-particle charge $\hat q_1$.

Several general features are common to all three observables. The extracted energy
first increases with increasing decay radius, reaches a broad maximum inside the
ergoregion, and subsequently decreases as the decay point approaches the outer
boundary of the ergoregion. The efficiencies exhibit the same qualitative behavior.
 In addition, all curves display a characteristic
sawtooth-like modulation. This structure is not numerical noise but a direct
consequence of the discreteness of the repetitive process: since the terminal
iteration number $n_f(\hat r)$ changes only by integer values while $\hat r$
varies continuously, every final-state quantity inherits the same staircase-like
oscillatory behavior, consistent with previous studies of the repetitive Penrose
process in accelerating Kerr spacetimes~\cite{ZengWang2026}.

Panel~(a) isolates the influence of the black-hole charge for fixed particle charges.
The case $\hat Q=-0.5$ consistently produces the smallest extracted energy and
efficiencies because the captured fragment experiences an electromagnetic repulsion
($\hat Q\hat q_1>0$), which weakens the negative-energy channel. For positively charged
black holes ($\hat Q>0$), the attractive coupling
($\hat Q\hat q_1<0$) enhances the negative-energy states and increases the extraction
efficiency. These results show that, once the particles carry electric charge, the sign
of the black-hole--particle interaction becomes more important than the magnitude of
$\hat Q$ itself.

Panel~(b) illustrates the role of the incident-particle charge. The extracted energy and
the EUE attain their largest values for a moderately positive charge
($\hat q_0\simeq0.2$), where the electrostatic repulsion remains weaker than the
gravitational attraction and the incident particle can repeatedly access the ergoregion.
As $\hat q_0$ increases further, the electrostatic barrier suppresses the number of
admissible extraction cycles, reducing both the total extracted energy and the EUE.
The EROI, however, increases because fewer but more energetic extraction events are
performed. For negative $\hat q_0$, the incident trajectory becomes less favorable for
repetitive extraction, leading to the lowest efficiencies among the cases considered.

Panel~(c) demonstrates that the captured-particle charge is the dominant parameter of
the repetitive Penrose process. Increasing the magnitude of a negatively charged captured
fragment progressively deepens the negative-energy states and enhances all three
performance measures. Conversely, a positively charged captured particle weakens the
negative-energy channel and produces the lowest extraction efficiencies. The trends
observed here for moderate values of $|\hat q_1|$ represent the onset of the much
stronger behavior discussed below, where large negative values of $\hat q_1$ drive the
system toward the near-reversible regime and allow the efficiency to exceed the
previously established limits for neutral Kerr and purely electric
Reissner--Nordström black holes.

Taken together, Figs.~\ref{fig:Eext}--\ref{fig:EROI} show that the terminal performance
of the repetitive Penrose process in Kerr--Newman spacetime is governed by the interplay
between two independent electromagnetic couplings. The coupling
$\hat{Q}\hat{q}_0$ controls the accessibility of the ergoregion by the incident particle
and therefore determines the number of admissible extraction cycles before the process
terminates. In contrast, the coupling $\hat{Q}\hat{q}_1$ determines the depth of the
negative-energy states available to the captured fragment and thus directly controls the
amount of energy extracted in each cycle. The highest overall efficiency is obtained when
the incident particle carries the same sign of charge as the black hole
($\hat{Q}\hat{q}_0>0$), ensuring efficient access to the ergoregion, while the captured
fragment carries the opposite sign ($\hat{Q}\hat{q}_1<0$), allowing increasingly
negative values of $\hat{E}_1$. This charge-sign selectivity has no analogue in either
the neutral Kerr or the purely charged Reissner--Nordström repetitive Penrose process
and constitutes one of the defining characteristics of energy extraction in the
Kerr--Newman geometry.

\newpage

\section{Results and Discussion}
\label{sec:results}
 
Having established the analytic solution of the conservation laws together with the
iterative evolution of $M_n$, $L_n$, $Q_n$, and $M_{{\rm irr},n}$ after each decay
event, we now present the numerical results. We begin by isolating the effect of the
black-hole charge through a comparison of neutral-particle extraction in Kerr and
charged backgrounds. We then investigate the influence of particle charges and identify
the electromagnetic couplings that govern the efficiency of the repetitive Penrose
process. Next, we examine the evolution toward large negative values of the
captured-particle charge, where a transient spin-up of the dimensionless spin parameter
emerges, and subsequently map the corresponding phase structure of the parameter space.
Finally, we analyze the nonlinear evolution of the black-hole charge and compare the
resulting behavior with previously studied repetitive Penrose processes in other
black-hole geometries.

\subsection{Neutral-particle baseline: Kerr versus Kerr--Newman}
\label{subsec:neutral_baseline}

We begin by establishing the neutral reference case through a comparison of the
repetitive Penrose process in Kerr ($\hat Q=0$) and Kerr--Newman
($\hat Q=0.5$) spacetimes, with electrically neutral particles
($\hat q_0=\hat q_1=0$). The corresponding numerical results are listed in
Tables~\ref{P0} and~\ref{P1}.

For identical initial conditions, the Kerr black hole consistently yields larger values
of both the energy return on investment (EROI), $\xi_n$, and the energy utilization
efficiency (EUE), $\Xi_n$, whereas the irreducible mass increases more rapidly in the
charged Kerr--Newman background. Although the presence of electric charge enlarges the
initial extractable energy reservoir of the black hole, a greater fraction of this
energy is irreversibly converted into the irreducible mass rather than being transferred
to the escaping particles.

This comparison shows that the electric charge of the black hole, by itself, does not
improve the efficiency of the repetitive Penrose process. Instead, in the absence of
charged particles, it enhances the irreversible growth of the irreducible mass and
thereby reduces the fraction of the available energy that can ultimately be extracted.
These results show that the black-hole charge alone cannot enhance the repetitive
Penrose process. The situation changes qualitatively, however, once the incident and
captured particles are allowed to carry electric charge, introducing additional
electromagnetic interactions that directly affect the extraction mechanism.

\subsection{Charged particles: electromagnetic coupling and charge-sign selectivity}
\label{subsec:charge_sign}

The situation changes once the incident and decay particles are allowed to carry
electric charge. Tables~\ref{P2}--\ref{P4} present representative results for a
Kerr--Newman black hole with $\hat Q=0.5$ and different combinations of
$\hat q_0$ and $\hat q_1$. In contrast to the neutral-particle case, both the
EROI and the EUE now depend strongly on the magnitude and, more importantly, the
relative signs of the particle charges.

The influence of the captured-particle charge is illustrated by comparing
Tables~\ref{P2} and~\ref{P4}, where the incident particle has the same charge
($\hat q_0=0.2$), while the captured fragment changes from
$\hat q_1=+5$ to $\hat q_1=-5$. Reversing the sign of the captured-particle charge
produces a substantial increase in the extracted energy and in both efficiency
measures. This enhancement originates from the attractive electromagnetic interaction
($\hat Q\hat q_1<0$), which allows the captured fragment to occupy more negative-energy
states inside the ergoregion and consequently transfers a larger fraction of the
black-hole rotational energy to the escaping particle.

The role of the incident-particle charge is seen by comparing
Tables~\ref{P3} and~\ref{P4}, where the captured-particle charge is fixed
($\hat q_1=-5$), while the incident-particle charge changes from
$\hat q_0=-0.2$ to $\hat q_0=0.2$. The positively charged incident particle leads to
systematically higher efficiencies. Since the incident particle reaches the ergoregion
from infinity, the quantity $\hat Q\hat q_0$ determines the strength of the
electromagnetic interaction during its approach to the black hole. For
$\hat Q\hat q_0>0$, the electrostatic interaction remains sufficiently weak compared
with gravity for the particle to access the ergoregion efficiently, allowing a larger
fraction of the available energy to participate in the repetitive Penrose process.

These comparisons show that the two electromagnetic couplings,
$\hat Q\hat q_0$ and $\hat Q\hat q_1$, govern different aspects of the extraction
mechanism. The former primarily controls the accessibility of the ergoregion by the
incident particle, whereas the latter determines the depth of the negative-energy
channel available to the captured fragment. Consequently, the most favorable
configuration is
\begin{equation}
\hat Q\hat q_0>0,
\qquad
\hat Q\hat q_1<0,
\end{equation}
corresponding to an incident particle carrying the same sign of charge as the black
hole and a captured fragment carrying the opposite sign. This charge-sign dependence
is a characteristic feature of the Kerr--Newman geometry and has no counterpart in the
neutral Kerr spacetime.

Motivated by these observations, the remainder of this work focuses exclusively on
this optimal configuration in order to determine how the efficiency evolves as the
magnitude of the captured-particle charge increases and to identify the physical
limits imposed by black-hole thermodynamics.

\subsection{Efficiency enhancement and transient spin-up}
\label{subsec:evolution}

Having identified the optimal charge configuration
\[
\hat Q\hat q_0>0,\qquad
\hat Q\hat q_1<0,
\]
we now investigate how the repetitive Penrose process evolves as the magnitude of the
captured-particle charge increases while the black-hole charge and the incident-particle
charge remain fixed ($\hat Q=0.5$, $\hat q_0=0.2$). The corresponding numerical results
are summarized in Tables~\ref{P4}--\ref{P-44.01}.

As $|\hat q_1|$ increases, the attractive electromagnetic interaction between the
positively charged black hole and the negatively charged captured fragment becomes
progressively stronger. Consequently, the captured particle reaches increasingly
negative-energy states inside the ergoregion, allowing a larger fraction of the
black-hole rotational energy to be transferred to the escaping particle. Both the
EROI and the EUE therefore increase monotonically throughout this parameter range. For
example, the final utilization efficiency rises from
$\Xi_n\simeq0.168$ at $\hat q_1=-5$
(Table~\ref{P4}) to
$\Xi_n\simeq0.317$ at $\hat q_1=-44.01$
(Table~\ref{P-44.01}), corresponding to an enhancement of almost a factor of two.

Besides improving the extraction efficiency, increasing $|\hat q_1|$ also modifies the
evolution of the black-hole spin. For relatively small values of $|\hat q_1|$, the
dimensionless spin parameter
$
\hat a_n=\frac{L_n}{M_n^{\,2}}
$
decreases monotonically, in agreement with the conventional picture of rotational
energy extraction. A qualitatively different behavior appears once the captured-particle
charge exceeds the threshold
$
|\hat q_1|\simeq16.07.
$

Beyond this point, the total angular momentum of the black hole continues to decrease
after each extraction event, but the mass decreases even more rapidly owing to the
strong attractive electromagnetic interaction. Since the dimensionless spin parameter
depends on the ratio $L/M^2$, the faster reduction of the mass temporarily dominates
over the loss of angular momentum, producing a transient increase of $\hat a_n$.

It is important to emphasize that this spin-up does not imply an increase in the total
angular momentum of the black hole. Instead, it results entirely from the different
rates at which the mass and angular momentum evolve during the repetitive process.
Throughout the evolution,
$
L_n
$
decreases monotonically, whereas
$
M_n^2
$
decreases even faster, causing their ratio to increase over a finite interval.

This behavior has no counterpart in the repetitive Penrose process in the neutral Kerr
geometry, where the evolution of the dimensionless spin remains strictly monotonic.
Here the electromagnetic interaction introduces an additional mechanism that modifies
the relative evolution of the conserved quantities, making the captured-particle charge
not only the principal parameter controlling the extraction efficiency but also the
quantity governing the dynamical evolution of the black-hole spin.

 %------------------------------------------------------------
\subsection{Phase structure of the parameter space}
\label{subsec:phase}
%------------------------------------------------------------

A qualitatively new picture emerges as the magnitude of the captured-particle charge is
increased beyond the moderate-charge regime discussed above. Instead of evolving
smoothly toward higher efficiencies, the repetitive Penrose process undergoes a sequence
of distinct dynamical transitions, indicating that the parameter space naturally
separates into four physically different regions. Their boundaries and physical
interpretation are summarized in Table~\ref{tab:q1_phase}.

The first transition occurs at approximately
\[
\hat q_1\simeq-44.02.
\]
For
$
-54.22\lesssim\hat q_1\lesssim-44.02,
$
the repetitive Penrose process no longer persists beyond the first decay event.
Although the initial fragmentation still produces a negative-energy captured particle,
the extraction reduces the black-hole spin below the minimum value required for the next
Penrose event,
$
\hat a_1<\hat a_{{\rm min},1},
$
and the iterative sequence therefore terminates immediately. The disappearance of the
repetitive process over a finite interval of $\hat q_1$ is a genuine nonlinear feature
of the coupled evolution of the black-hole mass, angular momentum, and electric charge,
and has no analogue in the neutral Kerr spacetime.

A second transition takes place when the captured-particle charge becomes slightly more
negative. For
\[
-54.85405\lesssim\hat q_1\lesssim-54.22,
\]
the repetitive Penrose process reappears and enters an exceptionally efficient regime.
The increasingly attractive electromagnetic interaction drives the captured fragment to
extremely negative-energy states, allowing a much larger fraction of the rotational
energy to be transferred to the escaping particle. As a result, both the EROI and the
EUE increase sharply, with the EUE approaching unity.

The lower boundary of this interval,
\[
\hat q_1\simeq-54.85405,
\]
defines a critical charge. As shown in Table~\ref{P-54.85405}, the change in the
irreducible mass during the first extraction cycles becomes negligibly small,
$
\Delta M_{\rm irr}\approx0,
$
indicating that the process approaches the reversible limit originally identified by
Christodoulou and Ruffini. In this regime almost all of the extracted energy originates
from the rotational and electromagnetic reservoirs, while only an infinitesimal fraction
is converted into irreducible mass. This critical configuration therefore represents the
highest physically admissible efficiency of the repetitive Penrose process in
Kerr--Newman spacetime.

The character of the evolution changes once the captured-particle charge exceeds this
critical value. For
\[
\hat q_1<-54.85405,
\]
negative-energy states remain kinematically accessible and the conservation laws admit
formal Penrose solutions. The thermodynamic behavior, however, becomes unphysical:
the irreducible mass decreases after every extraction event,
$
\Delta M_{\rm irr}<0,
$
in direct contradiction with Hawking's area theorem.

The appearance of this region therefore does not indicate a new physically realizable
mode of energy extraction. Instead, it marks the boundary beyond which the
test-particle approximation ceases to be self-consistent. Once the extracted energy
approaches the reversible limit, the neglected self-force, electromagnetic
backreaction, and finite-size effects can no longer be ignored. The critical value
$
\hat q_1\simeq-54.85405
$
thus represents the practical limit of the physically meaningful parameter space rather
than merely a kinematic boundary.

Taken together, these results reveal that the repetitive Penrose process in
Kerr--Newman spacetime is characterized by a four-region phase structure consisting of
an ordinary extraction regime, a disconnected forbidden interval, a second allowed
region with dramatically enhanced efficiency, and a thermodynamically forbidden region.
The present results therefore reveal a qualitatively richer dynamical structure than
that found in previously studied repetitive Penrose processes, including the Kerr,
Reissner--Nordström, Kerr--de~Sitter, and accelerating Kerr spacetimes.

\begin{table}[ht]
\centering
\caption{%
  Classification of the repetitive Penrose process as a function of the
  captured-particle charge $\hat{q}_1$, for fixed $\hat{Q} = 0.5$ and
  $\hat{q}_0 = 0.2$.
}
\label{tab:q1_phase}
\renewcommand{\arraystretch}{1.35}
\small
\setlength{\tabcolsep}{4pt}
\begin{tabular}{|c|c|p{3.2cm}|p{6.4cm}|}
\hline
\textbf{Range of $\hat{q}_1$}
  & \textbf{Region}
  & \textbf{Status}
  & \textbf{Physical interpretation} \\
\hline
$-44.02 < \hat{q}_1 \leq 0$ & I & Allowed &
  Ordinary repetitive extraction.
  $\Delta M_{\rm irr} > 0$; the spin condition
  $\hat{a}_n > \hat{a}_{{\rm min},n}$ is satisfied throughout;
  EROI and EUE increase gradually with $|\hat{q}_1|$. \\
\hline
$-54.22 < \hat{q}_1 \leq -44.02$ & II & Forbidden &
  The repetitive sequence terminates after the first decay because
  $\hat{a}_1 < \hat{a}_{{\rm min},1}$.
  Negative-energy states exist but the iterative process cannot continue
  (see Appendix~\ref{app:regionII}). \\
\hline
$-54.85405 \leq \hat{q}_1 \leq -54.22$ & III & Allowed &
  Repetitive process reappears with dramatically enhanced efficiency;
  EROI and EUE reach their maximum physically allowed values.
  Near the lower boundary, $\Delta M_{\rm irr} \approx 0$
  (near-reversible Christodoulou--Ruffini limit). \\
\hline
$\hat{q}_1 < -54.85405$ & IV & Thermodynamically forbidden &
  Negative-energy states remain kinematically accessible, but
  $\Delta M_{\rm irr} < 0$ violates
  Hawking's area theorem; physically inadmissible. \\
\hline
\end{tabular}
\end{table}

%------------------------------------------------------------
\subsection{Charge reversal and removal of the Reissner--Nordstr\"om barrier}
\label{subsec:charge_reversal}
%------------------------------------------------------------

\begin{figure}[htbp]
\centering
\includegraphics[width=\textwidth]{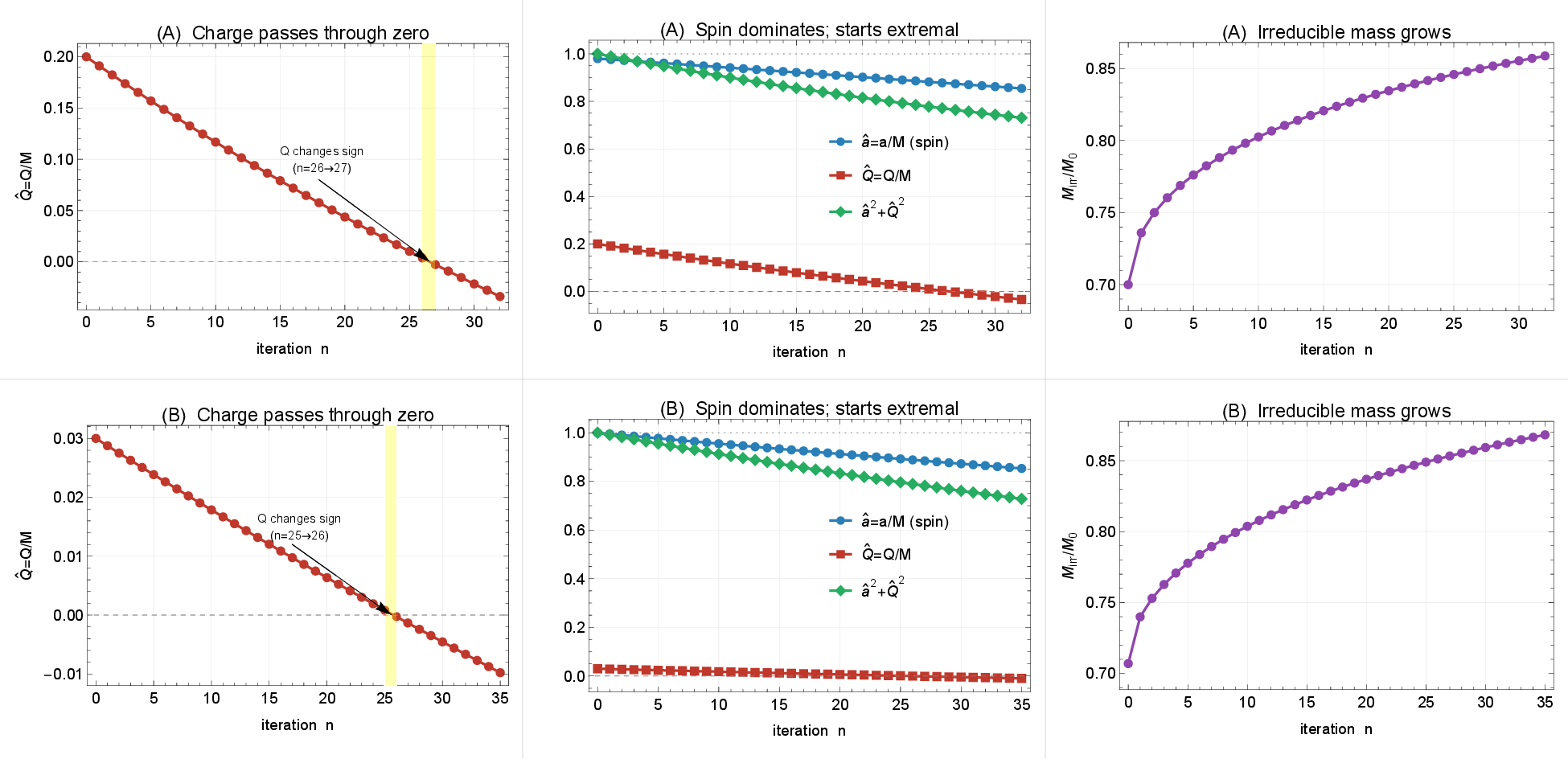}
\caption{Charge-sign reversal under the repetitive Penrose process in an initially
extremal Kerr--Newman black hole, shown for two spin-dominated configurations:
row~(A) a moderately charged hole ($\hat{Q}_0=0.2$, $\hat{q}_1=-30$,
Table~\ref{PQ0:0.2q1:-30}) and row~(B) a weakly charged hole ($\hat{Q}_0=0.03$,
$\hat{q}_1=-5$, Table~\ref{PQ0:0.03q1:-5}). \textbf{Left:} the dimensionless charge
$\hat{Q}_n$ decreases monotonically with the iteration index $n$, passes through the
neutral state, and reverses sign (shaded band: $n=26\!\to\!27$ in~(A) and
$n=25\!\to\!26$ in~(B)), so that an initially positively charged hole becomes negatively
charged. \textbf{Center:} the dimensionless spin $\hat{a}_n$ (blue) and charge
$\hat{Q}_n$ (red) together with the extremality function $\hat{a}_n^{2}+\hat{Q}_n^{2}$
(green); the latter starts at unity (extremal) and remains strictly below~$1$, so weak
cosmic censorship is preserved. \textbf{Right:} the irreducible mass $M_{\rm irr}/M_0$
grows monotonically throughout ($\Delta M_{\rm irr}>0$), consistent with Hawking's area
theorem. The full numerical sequences are listed in Tables~\ref{PQ0:0.2q1:-30} and
\ref{PQ0:0.03q1:-5} of Appendix~\ref{app:charge-reversal}. Fixed parameters in both rows:
$\hat{q}_0=0.2$, $\hat{E}_0=1$, $\hat{r}=1.9$, $\nu=0.78345$, $\hat{p}_{\phi1}=-19.434$,
$\mu_0=10^{-2}M_0$, and $M_0=1$.}
\label{fig:charge-reversal}
\end{figure}

A second distinctive feature of the repetitive Penrose process in Kerr--Newman
spacetime is the nonlinear evolution of the black-hole electric charge.
Throughout the iterative extraction, the charge is updated according to the charge
conservation law after each capture event. When the rotational contribution dominates
over the electromagnetic one ($\hat a>\hat Q$), the numerical evolution shows that the
dimensionless charge decreases continuously, reaches the neutral configuration,
$
\hat Q_n=0,
$
and subsequently changes sign while the process continues to satisfy
\[
\Delta M_{\rm irr}\ge0,
\qquad
\hat a_n^{\,2}+\hat Q_n^{\,2}<1.
\]
Representative numerical sequences are listed in
Tables~\ref{PQ0:0.2q1:-30} and~\ref{PQ0:0.03q1:-5} of
Appendix~\ref{app:charge-reversal}, and the corresponding evolution is illustrated in
Fig.~\ref{fig:charge-reversal}.

This behavior differs fundamentally from the repetitive Penrose process in the
Reissner--Nordström spacetime. In the purely charged case, the black-hole charge
approaches the neutral state asymptotically but never reaches it, leading to the
third-law-like statement that a charged black hole cannot be completely discharged
through a finite sequence of Penrose events. The presence of rotation removes this
restriction. The rotational energy reservoir provides an additional extraction channel
that allows the electric charge to cross the neutral state and evolve into the
opposite charge sector without violating either Hawking's area theorem or the
condition
$
\hat a_n^{\,2}+\hat Q_n^{\,2}<1.
$

The charge reversal therefore provides another manifestation of the nonlinear coupling
between rotation and electromagnetism in Kerr--Newman spacetime. Unlike the neutral
Kerr or purely charged Reissner--Nordström geometries, the simultaneous evolution of
the mass, angular momentum, and electric charge permits a qualitatively new mode of
black-hole evolution while remaining entirely consistent with black-hole
thermodynamics.

\newpage

\section{Conclusion}
\label{sec:conclusion}

In this work we have developed a repetitive Penrose process for charged particles in an
initially extremal Kerr--Newman black hole. The formalism provides a fully iterative
description of the extraction process, in which the black-hole mass, angular momentum,
electric charge, and irreducible mass are updated after every decay event. By combining
the conservation laws with the triple turning-point condition, we obtained a complete
analytic description of each extraction step and followed the evolution of the black hole
through successive Penrose decays.

The numerical analysis shows that the presence of electric charge changes the repetitive
Penrose process in several important ways. The coupling between the black-hole charge and
the incident particle, $\hat Q\hat q_0$, determines whether the particle can efficiently
reach the ergoregion and therefore controls the lifetime of the repetitive sequence. In
contrast, the coupling $\hat Q\hat q_1$ governs the depth of the negative-energy states
available to the captured fragment and consequently becomes the dominant parameter
controlling the extraction efficiency. The most favorable configuration is obtained when
the incident particle has the same sign as the black-hole charge while the captured
particle has the opposite sign, allowing both the energy return on investment and the
energy utilization efficiency to increase substantially.

As the magnitude of the captured-particle charge is increased, the repetitive Penrose
process undergoes a sequence of qualitative transitions. The parameter space separates
naturally into four distinct regions, including a narrow interval in which the evolution
approaches the reversible limit of black-hole thermodynamics,
$\Delta M_{\rm irr}\approx0$, and the energy utilization efficiency approaches unity.
Beyond the corresponding critical charge, the irreducible mass decreases and Hawking's
area theorem is violated, indicating that the test-particle approximation is no longer
physically reliable.

Another noteworthy result concerns the evolution of the black-hole electric charge.
Unlike the repetitive Penrose process in the Reissner--Nordström geometry, where the
black-hole charge approaches zero without changing sign, the Kerr--Newman black hole may
pass continuously through the neutral state and enter the opposite charge sector while
remaining consistent with both the area theorem and the extremality bound. The evolution
found here therefore indicates that the discharge barrier identified for purely charged
black holes is not a generic property of charged spacetimes, but instead results from the
absence of rotation.

Taken together, these results demonstrate that the simultaneous evolution of mass,
angular momentum, and electric charge produces extraction regimes that are absent when
rotation or electric charge acts independently. Charge-sign selectivity, temporary
spin-up, disconnected dynamical regions, operation close to the reversible limit, and
charge reversal all arise from the nonlinear coupling between the rotational and
electromagnetic degrees of freedom. These features distinguish the repetitive
Kerr--Newman Penrose process from all previously investigated repetitive extraction
mechanisms.

Table~\ref{tab:comparison} summarizes the principal differences between the repetitive
Penrose process in Kerr--Newman spacetime and those previously studied in Kerr,
Reissner--Nordström, Kerr--de~Sitter, and accelerating Kerr geometries.

\begin{table*}[htbp]
\centering
\caption{Comparison of the repetitive Penrose process across black-hole backgrounds.
$M_{\rm irr}$ denotes the irreducible mass; EUE is the energy utilization efficiency.}
\label{tab:comparison}
\renewcommand{\arraystretch}{1.3}
\small
\setlength{\tabcolsep}{4pt}
\resizebox{\textwidth}{!}{%
\begin{tabular}{|l|c|c|c|p{5.5cm}|}
\hline
\textbf{Background} & \textbf{Extra hair} & \textbf{Control parameter} &
\textbf{EUE $>50\%$?} & \textbf{Key physical mechanism} \\
\hline
Kerr~\cite{Ruffini2025PRR} & spin & decay radius $\hat r$ & No &
Rotational energy mostly converted into $M_{\rm irr}$; spin cannot be fully drained. \\
\hline
Reissner--Nordström~\cite{HuCaiWang2026} & charge & $\hat q_1$, $\hat r$ & No &
Charge cannot reach zero (third-law analogue); EUE bounded below $50\%$. \\
\hline
Kerr--de~Sitter~\cite{WangZeng2025} & spin, $\Lambda$ & $\Lambda M^2$, $\hat r$ & No &
A positive cosmological constant enhances the EROI but the EUE remains below $50\%$. \\
\hline
Accelerating Kerr~\cite{ZengWang2026} & spin, acceleration & $\hat A$, $\hat r$ & Yes &
At small $\hat{r}$, acceleration redirects a larger fraction of the extractable energy toward the escaping fragment, allowing  EUE $>50\%$. \\
\hline
\textbf{Kerr--Newman (this work)} &
\textbf{spin, charge} &
$\mathbf{\hat Q\hat q_0,\ \hat Q\hat q_1}$ &
\textbf{Yes} &
Simultaneous extraction of rotational and electromagnetic energy. Attractive electromagnetic coupling drives the evolution toward the reversible thermodynamic limit with EUE approaching unity, while the black-hole charge may pass through zero and reverse sign, demonstrating that the Reissner--Nordström third-law analogue is not universal. \\
\hline
\end{tabular}}
\end{table*}

Several directions remain open for future investigation. Incorporating
electromagnetic self-force and gravitational backreaction would clarify the
behavior of the near-critical regime beyond the test-particle approximation.
It would also be interesting to extend the present framework to
Kerr--Newman--(anti-)de~Sitter and accelerating Kerr--Newman spacetimes, where
rotation, electric charge, acceleration, and the cosmological constant coexist.
Finally, it would be of interest to establish the connection between the
particle-based Penrose process developed here and charged superradiant
scattering, and to determine whether the charge-sign selectivity and the
near-reversible regime identified in this work persist in the corresponding
wave description.

 \newpage 
  \appendix

\section{Explicit coefficients of the analytic solution}
\label{app:coefficients}

This appendix gives the explicit expressions for the coefficients
$A$, $B$, and $D_{\rm KN}$ appearing in
Eq.~(\ref{eq:mu1}).
  The coefficients are obtained by substituting the turning-point solutions for
$\hat{p}_{\phi0}$ and $\hat{E}_1$,
Eqs.~(\ref{eq:pphi0})--(\ref{eq:E1}),
into the conservation equations and solving the resulting quadratic equation for
$\tilde{\mu}_1$. They depend on the metric components
$g^{tt}$, $g^{t\phi}$, and $g^{\phi\phi}$ evaluated on the equatorial plane at the decay radius,
together with the black-hole parameters
$\hat a$, $\hat Q$,
and the particle parameters
$\hat E_0$, $\hat p_{\phi1}$,
$\hat q_0$, $\hat q_1$,
and $\nu=\mu_2/\mu_1$.

  \begin{align}
A ={} & \frac{\hat{a} \hat{E}_0 M g^{t\phi} (\hat{q}_1 \hat{Q})}{\hat{r}}
+ \frac{\hat{a} \hat{E}_1 M \hat{Q} g^{t\phi} (\hat{q}_0 \hat{Q})}{\hat{r}}
- \frac{\hat{a} M^2 g^{\phi\phi} \hat{p}_{\phi 0} (\hat{q}_1 \hat{Q})}{\hat{r}}  
- \frac{\hat{a} M^2 g^{\phi\phi} \hat{p}_{\phi 1} (\hat{q}_0 \hat{Q})}{\hat{r}}+ \frac{g^{tt} (\hat{q}_0 \hat{q}_1 \hat{Q}^2)}{\hat{r}^2}\notag \\
&
+ \frac{\hat{a}^2 M^2 g^{\phi\phi} (\hat{q}_0 \hat{q}_1 \hat{Q}^2)}{\hat{r}^2}
- \frac{2 \hat{a} M g^{t\phi} (\hat{q}_0 \hat{q}_1 \hat{Q}^2)}{\hat{r}^2}
- \hat{E}_1 M g^{t\phi} \hat{p}_{\phi 0}
- \hat{E}_0 M g^{t\phi} \hat{p}_{\phi 1}+ \frac{M g^{t\phi} \hat{p}_{\phi 1} (\hat{q}_0 \hat{Q})}{\hat{r}} \notag \\
& 
- \frac{\hat{E}_1 g^{tt} (\hat{q}_0 \hat{Q})}{\hat{r}}
 - \frac{\hat{E}_0 g^{tt} (\hat{q}_1 \hat{Q})}{\hat{r}}
+ \hat{E}_0 \hat{E}_1 g^{tt}
+ M^2 g^{\phi\phi} \hat{p}_{\phi 0} \hat{p}_{\phi 1} 
 + \frac{M g^{t\phi} \hat{p}_{\phi 0} (\hat{q}_1 \hat{Q})}{\hat{r}}, \\[10pt]
B ={} & \frac{2 \hat{a} M \hat{Q} (\hat{E}_1 \hat{q}_1) g^{t\phi}}{\hat{r}}
- \frac{2 \hat{a} M^2 g^{\phi\phi} \hat{p}_{\phi 1} (\hat{q}_1 \hat{Q})}{\hat{r}}
+ \frac{\hat{a}^2 M^2 g^{\phi\phi} (\hat{q}_1^2 \hat{Q}^2)}{\hat{r}^2}
+ \frac{2 M g^{t\phi} \hat{p}_{\phi 1} (\hat{q}_1 \hat{Q})}{\hat{r}} \notag \\
& - \frac{2 \hat{a} M g^{t\phi} (\hat{q}_1^2 \hat{Q}^2)}{\hat{r}^2}
- 2 \hat{E}_1 M g^{t\phi} \hat{p}_{\phi 1}
- \frac{2 \hat{Q} (\hat{E}_1 \hat{q}_1) g^{tt}}{\hat{r}} 
 + \hat{E}_1^2 g^{tt}
+ M^2 g^{\phi\phi} \hat{p}_{\phi 1}^2
 + \frac{g^{tt} (\hat{q}_1^2 \hat{Q}^2)}{\hat{r}^2}
+ \nu^2, \\[10pt]
D_{\text{KN}} ={} & -g^{\phi\phi} M^2 \hat{E}_1^2 \hat{p}_{\phi 0}^2 g^{tt}
+ \frac{2 g^{\phi\phi} M^2 (\hat{Q} \hat{q}_1) \hat{E}_1 \hat{p}_{\phi 0}^2 g^{tt}}{\hat{r}}
- g^{\phi\phi} M^2 \hat{E}_0^2 \hat{p}_{\phi 1}^2 g^{tt} \notag \\
& + \frac{2 g^{\phi\phi} M^2 (\hat{Q} \hat{q}_0) \hat{E}_0 \hat{p}_{\phi 1}^2 g^{tt}}{\hat{r}}
- \frac{2 g^{\phi\phi} M^2 \hat{a} \hat{E}_0 \hat{E}_1 \hat{p}_{\phi 1} (\hat{Q} \hat{q}_0) g^{tt}}{\hat{r}}
- \frac{2 g^{\phi\phi} M^2 \hat{E}_1 \hat{p}_{\phi 0} \hat{p}_{\phi 1} (\hat{Q} \hat{q}_0) g^{tt}}{\hat{r}} \notag \\
& - \frac{2 g^{\phi\phi} M^2 \hat{a} \hat{E}_0 \hat{E}_1 \hat{p}_{\phi 0} (\hat{Q} \hat{q}_1) g^{tt}}{\hat{r}}
- \frac{2 g^{\phi\phi} M^2 \hat{E}_0 \hat{p}_{\phi 0} \hat{p}_{\phi 1} (\hat{Q} \hat{q}_1) g^{tt}}{\hat{r}}
- \frac{\nu^2 (\hat{Q}^2 \hat{q}_0^2) g^{tt}}{\hat{r}^2} \notag \\
& - \frac{g^{\phi\phi} M^2 \hat{a}^2 \hat{E}_1^2 (\hat{Q}^2 \hat{q}_0^2) g^{tt}}{\hat{r}^2}
- \frac{g^{\phi\phi} M^2 \hat{p}_{\phi 1}^2 (\hat{Q}^2 \hat{q}_0^2) g^{tt}}{\hat{r}^2}
- \frac{2 g^{\phi\phi} M^2 \hat{a} \hat{E}_1 \hat{p}_{\phi 0} (\hat{Q}^2 \hat{q}_0 \hat{q}_1) g^{tt}}{\hat{r}^2} \notag \\
& - \frac{2 g^{\phi\phi} M^2 \hat{a} \hat{E}_0 \hat{p}_{\phi 1} (\hat{Q}^2 \hat{q}_0 \hat{q}_1) g^{tt}}{\hat{r}^2}
- \frac{g^{\phi\phi} M^2 \hat{a}^2 \hat{E}_0^2 (\hat{Q}^2 \hat{q}_1^2) g^{tt}}{\hat{r}^2}
- \frac{g^{\phi\phi} M^2 \hat{p}_{\phi 0}^2 (\hat{Q}^2 \hat{q}_1^2) g^{tt}}{\hat{r}^2} \notag \\
& + \frac{2 \nu^2 (\hat{Q} \hat{q}_0) \hat{E}_0 g^{tt}}{\hat{r}}
+ \frac{2 g^{\phi\phi} M^2 \hat{a}^2 (\hat{Q}^2 \hat{q}_0 \hat{q}_1) \hat{E}_0 \hat{E}_1 g^{tt}}{\hat{r}^2}
+ \frac{2 g^{\phi\phi} M^2 \hat{a} (\hat{Q} \hat{q}_0) \hat{E}_1^2 \hat{p}_{\phi 0} g^{tt}}{\hat{r}} \notag \\
& + \frac{2 g^{\phi\phi} M^2 \hat{a} (\hat{Q}^2 \hat{q}_1^2) \hat{E}_0 \hat{p}_{\phi 0} g^{tt}}{\hat{r}^2}
+ \frac{2 g^{\phi\phi} M^2 \hat{a} (\hat{Q} \hat{q}_1) \hat{E}_0^2 \hat{p}_{\phi 1} g^{tt}}{\hat{r}}
+ \frac{2 g^{\phi\phi} M^2 \hat{a} (\hat{Q}^2 \hat{q}_0^2) \hat{E}_1 \hat{p}_{\phi 1} g^{tt}}{\hat{r}^2} \notag \\
& + \frac{2 g^{\phi\phi} M^2 (\hat{Q}^2 \hat{q}_0 \hat{q}_1) \hat{p}_{\phi 0} \hat{p}_{\phi 1} g^{tt}}{\hat{r}^2}
+ 2 g^{\phi\phi} M^2 \hat{E}_0 \hat{E}_1 \hat{p}_{\phi 0} \hat{p}_{\phi 1} g^{tt}
- \frac{2 M \nu^2 \hat{a} \hat{E}_0 (\hat{Q} \hat{q}_0) g^{t\phi}}{\hat{r}} \notag \\
& - \frac{2 M \nu^2 \hat{p}_{\phi 0} (\hat{Q} \hat{q}_0) g^{t\phi}}{\hat{r}}
+ \frac{2 M \nu^2 \hat{a} (\hat{Q}^2 \hat{q}_0^2) g^{t\phi}}{\hat{r}^2}
+ 2 M \nu^2 \hat{E}_0 \hat{p}_{\phi 0} g^{t\phi} \notag \\
& + \frac{M^2 \hat{a}^2 (\hat{Q}^2 \hat{q}_0^2) \hat{E}_1^2 (g^{t\phi})^2}{\hat{r}^2}
+ \frac{2 M^2 \hat{a} (\hat{Q} \hat{q}_0) \hat{E}_0 \hat{E}_1 \hat{p}_{\phi 1} (g^{t\phi})^2}{\hat{r}}
+ \frac{2 M^2 (\hat{Q} \hat{q}_0) \hat{E}_1 \hat{p}_{\phi 0} \hat{p}_{\phi 1} (g^{t\phi})^2}{\hat{r}} \notag \\
& - M^2 \nu^2 \hat{p}_{\phi 0}^2 g^{\phi\phi}
- \frac{M^2 \nu^2 \hat{a}^2 (\hat{Q}^2 \hat{q}_0^2) g^{\phi\phi}}{\hat{r}^2}
+ \frac{2 M^2 \nu^2 \hat{a} (\hat{Q} \hat{q}_0) \hat{p}_{\phi 0} g^{\phi\phi}}{\hat{r}} \notag \\
& - g^{tt} \nu^2 \hat{E}_0^2
+ \frac{(g^{t\phi})^2 M^2 \hat{a}^2 (\hat{Q}^2 \hat{q}_1^2) \hat{E}_0^2}{\hat{r}^2}
+ (g^{t\phi})^2 M^2 \hat{E}_1^2 \hat{p}_{\phi 0}^2 \notag \\
& + \frac{(g^{t\phi})^2 M^2 (\hat{Q}^2 \hat{q}_1^2) \hat{p}_{\phi 0}^2}{\hat{r}^2}
+ (g^{t\phi})^2 M^2 \hat{E}_0^2 \hat{p}_{\phi 1}^2
+ \frac{(g^{t\phi})^2 M^2 (\hat{Q}^2 \hat{q}_0^2) \hat{p}_{\phi 1}^2}{\hat{r}^2} \notag \\
& - \frac{2 (g^{t\phi})^2 M^2 \hat{E}_0 \hat{p}_{\phi 1}^2 (\hat{Q} \hat{q}_0)}{\hat{r}}
- \frac{2 (g^{t\phi})^2 M^2 \hat{a} \hat{E}_1^2 \hat{p}_{\phi 0} (\hat{Q} \hat{q}_0)}{\hat{r}}
- \frac{2 (g^{t\phi})^2 M^2 \hat{E}_1 \hat{p}_{\phi 0}^2 (\hat{Q} \hat{q}_1)}{\hat{r}} \notag \\
& - \frac{2 (g^{t\phi})^2 M^2 \hat{a} \hat{E}_0^2 \hat{p}_{\phi 1} (\hat{Q} \hat{q}_1)}{\hat{r}}
- \frac{2 (g^{t\phi})^2 M^2 \hat{a} \hat{E}_1 \hat{p}_{\phi 1} (\hat{Q}^2 \hat{q}_0^2)}{\hat{r}^2}
- \frac{2 (g^{t\phi})^2 M^2 \hat{a}^2 \hat{E}_0 \hat{E}_1 (\hat{Q}^2 \hat{q}_0 \hat{q}_1)}{\hat{r}^2} \notag \\
& - \frac{2 (g^{t\phi})^2 M^2 \hat{p}_{\phi 0} \hat{p}_{\phi 1} (\hat{Q}^2 \hat{q}_0 \hat{q}_1)}{\hat{r}^2}
- \frac{2 (g^{t\phi})^2 M^2 \hat{a} \hat{E}_0 \hat{p}_{\phi 0} (\hat{Q}^2 \hat{q}_1^2)}{\hat{r}^2}
+ \frac{2 (g^{t\phi})^2 M^2 \hat{a} (\hat{Q}^2 \hat{q}_0 \hat{q}_1) \hat{E}_1 \hat{p}_{\phi 0}}{\hat{r}^2} \notag \\
& + \frac{2 (g^{t\phi})^2 M^2 \hat{a} (\hat{Q} \hat{q}_1) \hat{E}_0 \hat{E}_1 \hat{p}_{\phi 0}}{\hat{r}}
+ \frac{2 (g^{t\phi})^2 M^2 \hat{a} (\hat{Q}^2 \hat{q}_0 \hat{q}_1) \hat{E}_0 \hat{p}_{\phi 1}}{\hat{r}^2}
+ \frac{2 (g^{t\phi})^2 M^2 (\hat{Q} \hat{q}_1) \hat{E}_0 \hat{p}_{\phi 0} \hat{p}_{\phi 1}}{\hat{r}} \notag \\
& - 2 (g^{t\phi})^2 M^2 \hat{E}_0 \hat{E}_1 \hat{p}_{\phi 0} \hat{p}_{\phi 1}.
\end{align}

\section{Iterative data for the forbidden and critical charge windows}
\label{app:regionII}

This appendix presents representative numerical sequences for the charge intervals
discussed in Sec.~\ref{sec:results}. Throughout this appendix we fix
$\hat{E}_0=1$, $\hat{r}=1.2$, $\nu=0.78345$,
$\hat{p}_{\phi1}=-19.434$, $\mu_0=10^{-2}M_0$,
$\hat{Q}_0=0.5$, and $\hat{q}_0=0.2$, varying only the captured-particle charge
$\hat{q}_1$.

The tables for
$-54.22\lesssim\hat{q}_1\lesssim-44.02$
show that the repetitive process terminates after the first decay because
$\hat{a}_1<\hat{a}_{\rm min,1}$.
For
$-54.85405<\hat{q}_1\le-54.22$,
the process becomes allowed again and approaches the reversible limit,
$\Delta M_{\rm irr}\rightarrow0^{+}$.
For
$\hat{q}_1<-54.85405$,
the irreducible mass decreases,
$\Delta M_{\rm irr}<0$,
indicating the thermodynamically forbidden regime.
  
\begin{table}[htbp]
    \centering
    \setlength{\arrayrulewidth}{0.5pt}
    \renewcommand{\arraystretch}{1.2}
    \scriptsize
    \setlength{\tabcolsep}{3pt}
\resizebox{\textwidth}{!}{%
\begin{tabular}{ccccccccccccc}
\toprule
$n$ & $M/M_0$ & $\hat{a}_n$ & $\hat{Q}_n$ & $E_\text{extr}/M_0$ & $E_\text{ext}/M_0$ & $M_\text{irr}/M_0$ & $\tilde{\mu}_{1,n}$ & $\hat{E}_{1,n}$ & $\hat{p}_{\phi0,n}$ & $\hat{a}_{\min,0,n}$ & $\xi_n$ & $\Xi_n$\\
\midrule
 0 & 1 & 0.866025 & 0.5 & 0.338562 & 0 & 0.661438 & 0.029322 & -12.3598 & 2.007715 & 0.857057&  0 & 0 \\
 1 & 0.996376 & 0.866597 & 0.494136 & 0.310581 & 0.00362413 & 0.685795 & 0.0289261 & -12.3447 & 2.0298 & 0.860407 & 0.362413 & 0.129521 \\
 2 & 0.992805 & 0.867159 & 0.48835 & 0.298383 & 0.00719498 & 0.694422 & 0.0285272 & -12.333 & 2.05308 & 0.863661 & 0.359749 & 0.179072 \\
 3 & 0.989287 & 0.867714 & 0.482645 & 0.288924 & 0.0107132 & 0.700363 & 0.0281231 & -12.3249 & 2.07783 & 0.866822 & 0.357108 & 0.215825 \\
 \textcolor{red}{4} & \textcolor{red}{0.985821} & \textcolor{red}{0.868263} & \textcolor{red}{0.47702} & \textcolor{red}{0.280942} & \textcolor{red}{0.0141794} & \textcolor{red}{0.704879} & \textcolor{red}{0.0277105} & \textcolor{red}{-12.3213} & \textcolor{red}{2.10443} & \textcolor{red}{0.869891} & \textcolor{red}{0.354484} & \textcolor{red}{0.246082} \\
\bottomrule
\end{tabular}
}% end resizebox
\caption{Repetitive Penrose process at $\hat{q}_1=-20$ (Region~I), illustrating the
onset of the transient spin-up that sets in beyond $|\hat{q}_1|\simeq16.07$;
$\hat{q}_0=0.2$, $\hat{Q}_0=0.5$. Fixed parameters as in Table~\ref{P0}.}
\label{P-20}
\end{table}

\begin{table}[htbp]
    \centering
    \setlength{\arrayrulewidth}{0.5pt}
    \renewcommand{\arraystretch}{1.2}
    \scriptsize
    \setlength{\tabcolsep}{3pt}
\resizebox{\textwidth}{!}{%
\begin{tabular}{ccccccccccccc}
\toprule
$n$ & $M/M_0$ & $\hat{a}_n$ & $\hat{Q}_n$ & $E_\text{extr}/M_0$ & $E_\text{ext}/M_0$ & $M_\text{irr}/M_0$ & $\tilde{\mu}_{1,n}$ & $\hat{E}_{1,n}$ & $\hat{p}_{\phi0,n}$ & $\hat{a}_{\min,0,n}$ & $\xi_n$ & $\Xi_n$\\
\midrule
 0 & 1 & 0.866025 & 0.5 & 0.338562 & 0 & 0.661438 & 0.095081 &-19.4874 & 2.007715 & 0.857057&  0 & 0 \\
 \textcolor{red}{1} & \textcolor{red}{0.981471} & \textcolor{red}{0.87985} & \textcolor{red}{0.458145} & \textcolor{red}{0.280077} & \textcolor{red}{0.0185288} & \textcolor{red}{0.701394} & \textcolor{red}{0.0736172} & \textcolor{red}{-18.4113} & \textcolor{red}{2.0911} & \textcolor{red}{0.879855} & \textcolor{red}{1.85288} & \textcolor{red}{0.316811} \\
\bottomrule
\end{tabular}
}% end resizebox
\caption{Repetitive Penrose process at $\hat{q}_1=-44.02$ (Region~II, forbidden): the
process halts after the first decay because $\hat{a}_1<\hat{a}_{\mathrm{min},1}$;
$\hat{q}_0=0.2$, $\hat{Q}_0=0.5$. Fixed parameters as in Table~\ref{P0}.}
\label{P-44.02}
\end{table}

\begin{table}[htbp]
    \centering
    \setlength{\arrayrulewidth}{0.5pt}
    \renewcommand{\arraystretch}{1.2}
    \scriptsize
    \setlength{\tabcolsep}{3pt}
\resizebox{\textwidth}{!}{%
\begin{tabular}{ccccccccccccc}
\toprule
$n$ & $M/M_0$ & $\hat{a}_n$ & $\hat{Q}_n$ & $E_\text{extr}/M_0$ & $E_\text{ext}/M_0$ & $M_\text{irr}/M_0$ & $\tilde{\mu}_{1,n}$ & $\hat{E}_{1,n}$ & $\hat{p}_{\phi0,n}$ & $\hat{a}_{\min,0,n}$ & $\xi_n$ & $\Xi_n$\\
\midrule
 0 & 1 & 0.866025 & 0.5 & 0.338562 & 0 & 0.661438 & 0.37999 & -22.4275 & 2.007715 & 0.857057&  0 & 0 \\
 \textcolor{red}{1} & \textcolor{red}{0.914778} & \textcolor{red}{0.946655} & \textcolor{red}{0.294007} & \textcolor{red}{0.239853} & \textcolor{red}{0.0852224} & \textcolor{red}{0.674924} & \textcolor{red}{0.0546863} & \textcolor{red}{-15.7614} & \textcolor{red}{2.12608} & \textcolor{red}{0.946702} & \textcolor{red}{8.52224} & \textcolor{red}{0.863372} \\
\bottomrule
\end{tabular}
}% end resizebox
\caption{Repetitive Penrose process at $\hat{q}_1=-54.21$ (Region~II, forbidden):
representative case terminating at $n=0$ for the same reason as Table~\ref{P-44.02};
$\hat{q}_0=0.2$, $\hat{Q}_0=0.5$. Fixed parameters as in Table~\ref{P0}.}
\label{P-54.21}
\end{table}

\begin{table}[htbp]
    \centering
    \setlength{\arrayrulewidth}{0.5pt}
    \renewcommand{\arraystretch}{1.2}
    \scriptsize
    \setlength{\tabcolsep}{3pt}
\resizebox{\textwidth}{!}{%
\begin{tabular}{ccccccccccccc}
\toprule
$n$ & $M/M_0$ & $\hat{a}_n$ & $\hat{Q}_n$ & $E_\text{extr}/M_0$ & $E_\text{ext}/M_0$ & $M_\text{irr}/M_0$ & $\tilde{\mu}_{1,n}$ & $\hat{E}_{1,n}$ & $\hat{p}_{\phi0,n}$ & $\hat{a}_{\min,0,n}$ & $\xi_n$ & $\Xi_n$ \\
\midrule
 0 & 1 & 0.866025 & 0.5 & 0.338562 & 0 & 0.661438 & 0.404551 & -22.5423 & 2.007715 & 0.857057 & 0 & 0\\
 1 & 0.908805 & 0.95336 & 0.278994 & 0.242128 & 0.0911952 & 0.666677 & 0.0520852 & -15.3285 & 2.10798 & 0.95119 & 9.11952 & 0.945676 \\
 \textcolor{red}{2} & \textcolor{red}{0.900821} & \textcolor{red}{0.958998} & \textcolor{red}{0.25054} & \textcolor{red}{0.232419} & \textcolor{red}{0.0991791} & \textcolor{red}{0.668402} & \textcolor{red}{0.0450025} & \textcolor{red}{-14.4994} & \textcolor{red}{2.13534} & \textcolor{red}{0.959001} & \textcolor{red}{4.95895} & \textcolor{red}{0.934391} \\
\bottomrule
\end{tabular}
}% end resizebox
\caption{Repetitive Penrose process at $\hat{q}_1=-54.63$ (Region~III), approaching the
reversible limit with $\Xi$ close to unity and $\Delta M_{\mathrm{irr}}\to0^{+}$;
$\hat{q}_0=0.2$, $\hat{Q}_0=0.5$. Fixed parameters as in Table~\ref{P0}.}
\label{P-54.63}
\end{table}

\begin{table}[htbp]
    \centering
    \setlength{\arrayrulewidth}{0.5pt}
    \renewcommand{\arraystretch}{1.2}
    \scriptsize
    \setlength{\tabcolsep}{3pt}
\resizebox{\textwidth}{!}{%
\begin{tabular}{ccccccccccccc}
\toprule
$n$ & $M/M_0$ & $\hat{a}_n$ & $\hat{Q}_n$ & $E_\text{extr}/M_0$ & $E_\text{ext}/M_0$ & $M_\text{irr}/M_0$ & $\tilde{\mu}_{1,n}$ & $\hat{E}_{1,n}$ & $\hat{p}_{\phi0,n}$ & $\hat{a}_{\min,0,n}$ & $\xi_n$ & $\Xi_n$ \\
\midrule
 0 & 1 & 0.866025 & 0.5 & 0.338562 & 0 & 0.661438 & 0.431716 & -22.6699 & 2.007715 & 0.857057 & 0 & 0\\
 1 & 0.902131 & 0.961032 & 0.262125 & 0.247405 & 0.0978695 & 0.654726 & 0.0494264 & -14.8303 & 2.08635 & 0.955929 & 9.78695 & 1.07363 \\
 2 & 0.8948 & 0.966019 & 0.234891 & 0.237176 & 0.1052 & 0.657624 & 0.0432116 & -14.029 & 2.10934 & 0.962915 & 5.25998 & 1.03761 \\
 3 & 0.888738 & 0.969729 & 0.211081 & 0.229487 & 0.111262 & 0.659251 & 0.0387961 & -13.3422 & 2.13059 & 0.968361 & 3.70872 & 1.02005 \\
 \textcolor{red}{4} & \textcolor{red}{0.883562} & \textcolor{red}{0.972541} & \textcolor{red}{0.189705} & \textcolor{red}{0.223303} & \textcolor{red}{0.116438} & \textcolor{red}{0.660259} & \textcolor{red}{0.0354386} & \textcolor{red}{-12.7367} & \textcolor{red}{2.15091} & \textcolor{red}{0.972733} & \textcolor{red}{2.91095} & \textcolor{red}{1.01023} \\
\bottomrule
\end{tabular}
}% end resizebox
\caption{Repetitive Penrose process at $\hat{q}_1=-55.1$ (Region~IV,
thermodynamically forbidden): a further example of decreasing irreducible mass
($\Delta M_{\mathrm{irr}}<0$); $\hat{q}_0=0.2$, $\hat{Q}_0=0.5$. Fixed parameters as in
Table~\ref{P0}.}
\label{P-55.1}
\end{table}

\section{Charge-sign reversal sequences}
\label{app:charge-reversal}

This appendix presents the numerical sequences corresponding to the charge reversal
discussed in Sec.~\ref{sec:results} and illustrated in
Fig.~\ref{fig:charge-reversal}. In all cases the black hole is initially extremal,
$\hat a_0^{\,2}+\hat Q_0^{\,2}=1$, with rotation dominating over the electric charge.
During the repetitive Penrose process, the dimensionless charge $\hat Q_n$
decreases, crosses zero, and changes sign, while the irreducible mass increases
($\Delta M_{\rm irr}>0$) and the condition
$\hat a_n^{\,2}+\hat Q_n^{\,2}<1$
remains satisfied throughout the evolution.

\begin{table}[htbp]
    \centering
    \setlength{\arrayrulewidth}{0.5pt}
    \renewcommand{\arraystretch}{1.2}
    \scriptsize
    \setlength{\tabcolsep}{3pt}
\resizebox{\textwidth}{!}{%
\begin{tabular}{ccccccccccccc}
\toprule
$n$ & $M/M_0$ & $\hat{a}_n$ & $\hat{Q}_n$ & $E_\text{extr}/M_0$ & $E_\text{ext}/M_0$ & $M_\text{irr}/M_0$ & $\tilde{\mu}_{1,n}$ & $\hat{E}_{1,n}$ & $\hat{p}_{\phi0,n}$ & $\hat{a}_{\min,0,n}$ & $\xi_n$ & $\Xi_n$\\
\midrule
 0 & 1        & 0.979796 & 0.2 & 0.3 & 0 & 0.7 & 0.029528 & -3.48402 & 2.40499 & 0.836625&  0 & 0 \\
 1 & 0.998971 & 0.976065 & 0.191142 & 0.263022 & 0.00102876 & 0.73595 & 0.0290954 & -3.35773 & 2.4164 & 0.838389 & 0.102876 & 0.0278207 \\
 2 & 0.997994 & 0.972305 & 0.182413 & 0.248012 & 0.00200571 & 0.749982 & 0.0286761 & -3.2332 & 2.4278 & 0.840047 & 0.100285 & 0.0385804 \\
 3 & 0.997067 & 0.96852 & 0.17381 & 0.236707 & 0.00293286 & 0.76036 & 0.0282695 & -3.11035 & 2.43918 & 0.841602 & 0.0977621 & 0.0463381 \\
 4 & 0.996188 & 0.964711 & 0.165329 & 0.227355 & 0.00381214 & 0.768832 & 0.027875 & -2.98914 & 2.45055 & 0.84306 & 0.0953036 & 0.0524767 \\
 5 & 0.995355 & 0.960879 & 0.156967 & 0.219273 & 0.00464537 & 0.776081 & 0.027492 & -2.8695 & 2.46192 & 0.844424 & 0.0929074 & 0.0575444 \\
 6 & 0.994566 & 0.957028 & 0.148719 & 0.212108 & 0.00543425 & 0.782458 & 0.02712 & -2.75139 & 2.47328 & 0.845699 & 0.0905709 & 0.0618284 \\
 7 & 0.99382 & 0.953158 & 0.140583 & 0.205647 & 0.00618043 & 0.788173 & 0.0267585 & -2.63474 & 2.48463 & 0.846888 & 0.0882919 & 0.065503 \\
 8 & 0.993115 & 0.949272 & 0.132556 & 0.199751 & 0.00688545 & 0.793363 & 0.026407 & -2.51952 & 2.49599 & 0.847993 & 0.0860681 & 0.0686838 \\
 9 & 0.992449 & 0.945371 & 0.124633 & 0.194325 & 0.00755078 & 0.798124 & 0.0260652 & -2.40567 & 2.50735 & 0.84902 & 0.0838975 & 0.0714527 \\
 10 & 0.991822 & 0.941456 & 0.116814 & 0.189295 & 0.00817782 & 0.802527 & 0.0257324 & -2.29315 & 2.51871 & 0.84997 & 0.0817782 & 0.0738704 \\
 11 & 0.991232 & 0.937529 & 0.109094 & 0.184608 & 0.0087679 & 0.806624 & 0.0254085 & -2.18192 & 2.53007 & 0.850847 & 0.0797082 & 0.0759834 \\
 12 & 0.990678 & 0.933592 & 0.101472 & 0.18022 & 0.00932229 & 0.810458 & 0.025093 & -2.07193 & 2.54145 & 0.851654 & 0.0776858 & 0.0778285 \\
 13 & 0.990158 & 0.929645 & 0.0939437 & 0.176098 & 0.00984221 & 0.81406 & 0.0247856 & -1.96316 & 2.55283 & 0.852392 & 0.0757093 & 0.0794353 \\
 14 & 0.989671 & 0.92569 & 0.0865081 & 0.172213 & 0.0103288 & 0.817458 & 0.024486 & -1.85555 & 2.56421 & 0.853065 & 0.073777 & 0.0808282 \\
 15 & 0.989217 & 0.921727 & 0.0791623 & 0.168542 & 0.0107831 & 0.820675 & 0.0241938 & -1.74908 & 2.57561 & 0.853675 & 0.0718876 & 0.0820275 \\
 16 & 0.988794 & 0.917759 & 0.0719041 & 0.165066 & 0.0112063 & 0.823728 & 0.0239088 & -1.64371 & 2.58702 & 0.854225 & 0.0700394 & 0.0830504 \\
 17 & 0.988401 & 0.913786 & 0.0647315 & 0.161768 & 0.0115993 & 0.826633 & 0.0236307 & -1.53941 & 2.59844 & 0.854715 & 0.0682312 & 0.0839116 \\
 18 & 0.988037 & 0.90981 & 0.0576423 & 0.158633 & 0.0119631 & 0.829404 & 0.0233592 & -1.43614 & 2.60988 & 0.855149 & 0.0664615 & 0.0846239 \\
 19 & 0.987701 & 0.90583 & 0.0506345 & 0.155648 & 0.0122985 & 0.832053 & 0.0230941 & -1.33388 & 2.62132 & 0.855529 & 0.0647292 & 0.0851983 \\
 20 & 0.987393 & 0.901849 & 0.0437063 & 0.152803 & 0.0126066 & 0.83459 & 0.0228352 & -1.2326 & 2.63279 & 0.855856 & 0.0630329 & 0.0856446 \\
 21 & 0.987112 & 0.897866 & 0.0368557 & 0.150089 & 0.0128881 & 0.837023 & 0.0225823 & -1.13226 & 2.64426 & 0.856131 & 0.0613717 & 0.0859712 \\
 22 & 0.986856 & 0.893883 & 0.030081 & 0.147495 & 0.0131437 & 0.839361 & 0.022335 & -1.03284 & 2.65576 & 0.856358 & 0.0597443 & 0.0861859 \\
 23 & 0.986626 & 0.889901 & 0.0233805 & 0.145016 & 0.0133744 & 0.84161 & 0.0220934 & -0.934319 & 2.66726 & 0.856537 & 0.0581497 & 0.0862954 \\
 24 & 0.986419 & 0.88592 & 0.0167525 & 0.142643 & 0.0135809 & 0.843777 & 0.021857 & -0.836664 & 2.67879 & 0.85667 & 0.0565869 & 0.0863058 \\
 25 & 0.986236 & 0.88194 & 0.0101954 & 0.14037 & 0.0137637 & 0.845866 & 0.0216258 & -0.739854 & 2.69033 & 0.856758 & 0.0550549 & 0.0862227 \\
 26 & 0.986076 & 0.877964 & 0.00370763 & 0.138192 & 0.0139237 & 0.847884 & 0.0213997 & -0.643863 & 2.70189 & 0.856803 & 0.0535528 & 0.0860511 \\
 27 & 0.985938 & 0.87399 & -0.00271227 & 0.136104 & 0.0140615 & 0.849834 & 0.0211783 & -0.54867 & 2.71346 & 0.856806 & 0.0520797 & 0.0857954 \\
 28 & 0.985822 & 0.870021 & -0.00906577 & 0.134101 & 0.0141777 & 0.851721 & 0.0209617 & -0.454253 & 2.72505 & 0.856769 & 0.0506347 & 0.0854598 \\
 29 & 0.985727 & 0.866056 & -0.0153543 & 0.132178 & 0.0142729 & 0.853549 & 0.0207495 & -0.36059 & 2.73666 & 0.856692 & 0.049217 & 0.0850481 \\
 30 & 0.985652 & 0.862096 & -0.0215791 & 0.130332 & 0.0143477 & 0.85532 & 0.0205418 & -0.267659 & 2.74828 & 0.856577 & 0.0478258 & 0.0845637 \\
 31 & 0.985597 & 0.858142 & -0.0277417 & 0.128559 & 0.0144027 & 0.857039 & 0.0203383 & -0.175442 & 2.75992 & 0.856426 & 0.0464604 & 0.0840097 \\
 \textcolor{red}{32} & \textcolor{red}{0.985562} & \textcolor{red}{0.854193} & \textcolor{red}{-0.0338432} & \textcolor{red}{0.126855} & \textcolor{red}{0.0144384} & \textcolor{red}{0.858707} & \textcolor{red}{0.020139} & \textcolor{red}{-0.0839191} & \textcolor{red}{2.77158} & \textcolor{red}{0.856238} & \textcolor{red}{0.04512} & \textcolor{red}{0.0833891} \\
\bottomrule
\end{tabular}
}% end resizebox
\caption{Complete iterative sequence underlying the charge-sign reversal of
Fig.~\ref{fig:charge-reversal} (row~A), for a moderately charged, initially extremal
Kerr--Newman black hole with captured charge $\hat{q}_1=-30$. The dimensionless charge
$\hat{Q}_n$ decreases monotonically, passes through the neutral state, and reverses sign.
Fixed parameters: $\hat{q}_0=0.2$, $\hat{E}_0=1$, $\hat{r}=1.9$, $\nu=0.78345$,
$\hat{p}_{\phi1}=-19.434$, $\mu_0=0.01\,M_0$, $\hat{Q}_0=0.2$, and $M_0=1$.}
\label{PQ0:0.2q1:-30}
\end{table}

\begin{table}[htbp]
    \centering
    \setlength{\arrayrulewidth}{0.5pt}
    \renewcommand{\arraystretch}{1.2}
    \scriptsize
    \setlength{\tabcolsep}{3pt}
\resizebox{\textwidth}{!}{%
\begin{tabular}{ccccccccccccc}
\toprule
$n$ & $M/M_0$ & $\hat{a}_n$ & $\hat{Q}_n$ & $E_\text{extr}/M_0$ & $E_\text{ext}/M_0$ & $M_\text{irr}/M_0$ & $\tilde{\mu}_{1,n}$ & $\hat{E}_{1,n}$ & $\hat{p}_{\phi0,n}$ & $\hat{a}_{\min,0,n}$ & $\xi_n$ & $\Xi_n$\\
\midrule
 0 & 1        & 0.99955 & 0.03       & 0.293052      & 0           & 0.706948      & 0.0248638   & -0.575206 & 2.44837 & 0.85636&  0 & 0 \\
 1 & 0.999857 & 0.995002 & 0.0287568 & 0.259961 & 0.000143018 & 0.739896 & 0.0247308 & -0.575006 & 2.45632 & 0.856397 & 0.0143018 & 0.00432188 \\
 2 & 0.999715 & 0.990477 & 0.0275203 & 0.246763 & 0.000285222 & 0.752951 & 0.0245984 & -0.57483 & 2.46433 & 0.856432 & 0.0142611 & 0.00616176 \\
 3 & 0.999573 & 0.985974 & 0.0262903 & 0.236856 & 0.000426621 & 0.762717 & 0.0244668 & -0.574678 & 2.47242 & 0.856465 & 0.0142207 & 0.00759168 \\
 4 & 0.999433 & 0.981494 & 0.025067 & 0.228659 & 0.000567226 & 0.770774 & 0.0243358 & -0.574551 & 2.48057 & 0.856496 & 0.0141807 & 0.00880879 \\
 5 & 0.999293 & 0.977035 & 0.0238502 & 0.221558 & 0.000707048 & 0.777735 & 0.0242056 & -0.574448 & 2.48879 & 0.856526 & 0.014141 & 0.00988955 \\
 6 & 0.999154 & 0.972598 & 0.0226399 & 0.215238 & 0.000846097 & 0.783916 & 0.024076 & -0.574369 & 2.49708 & 0.856554 & 0.0141016 & 0.0108732 \\
 7 & 0.999016 & 0.968183 & 0.0214361 & 0.209511 & 0.000984382 & 0.789505 & 0.0239472 & -0.574315 & 2.50543 & 0.85658 & 0.0140626 & 0.0117832 \\
 8 & 0.998878 & 0.96379 & 0.0202388 & 0.204255 & 0.00112191 & 0.794623 & 0.023819 & -0.574285 & 2.51386 & 0.856605 & 0.0140239 & 0.0126346 \\
 9 & 0.998741 & 0.959419 & 0.0190478 & 0.199386 & 0.0012587 & 0.799355 & 0.0236914 & -0.574279 & 2.52236 & 0.856629 & 0.0139856 & 0.0134382 \\
 10 & 0.998605 & 0.955069 & 0.0178633 & 0.194841 & 0.00139476 & 0.803764 & 0.0235646 & -0.574298 & 2.53093 & 0.85665 & 0.0139476 & 0.0142016 \\
 11 & 0.99847 & 0.95074 & 0.016685 & 0.190573 & 0.00153009 & 0.807897 & 0.0234384 & -0.574342 & 2.53956 & 0.856671 & 0.0139099 & 0.0149307 \\
 12 & 0.998335 & 0.946434 & 0.0155131 & 0.186545 & 0.00166471 & 0.81179 & 0.0233129 & -0.57441 & 2.54827 & 0.85669 & 0.0138725 & 0.01563 \\
 13 & 0.998201 & 0.942148 & 0.0143475 & 0.182729 & 0.00179862 & 0.815472 & 0.023188 & -0.574503 & 2.55705 & 0.856707 & 0.0138355 & 0.0163032 \\
 14 & 0.998068 & 0.937884 & 0.0131881 & 0.179101 & 0.00193183 & 0.818967 & 0.0230638 & -0.57462 & 2.56591 & 0.856723 & 0.0137988 & 0.0169532 \\
 15 & 0.997936 & 0.933641 & 0.0120349 & 0.175642 & 0.00206436 & 0.822294 & 0.0229402 & -0.574762 & 2.57483 & 0.856737 & 0.0137624 & 0.0175824 \\
 16 & 0.997804 & 0.929419 & 0.0108879 & 0.172334 & 0.00219621 & 0.82547 & 0.0228173 & -0.574929 & 2.58383 & 0.856751 & 0.0137263 & 0.0181929 \\
 17 & 0.997673 & 0.925218 & 0.00974699 & 0.169165 & 0.0023274 & 0.828508 & 0.0226951 & -0.57512 & 2.5929 & 0.856762 & 0.0136906 & 0.0187864 \\
 18 & 0.997542 & 0.921039 & 0.00861224 & 0.166122 & 0.00245792 & 0.83142 & 0.0225734 & -0.575336 & 2.60205 & 0.856773 & 0.0136551 & 0.0193644 \\
 19 & 0.997412 & 0.91688 & 0.00748357 & 0.163196 & 0.00258779 & 0.834216 & 0.0224524 & -0.575578 & 2.61127 & 0.856782 & 0.01362 & 0.0199281 \\
 20 & 0.997283 & 0.912741 & 0.00636095 & 0.160377 & 0.00271702 & 0.836906 & 0.022332 & -0.575844 & 2.62056 & 0.85679 & 0.0135851 & 0.0204787 \\
 21 & 0.997154 & 0.908624 & 0.00524435 & 0.157657 & 0.00284562 & 0.839498 & 0.0222123 & -0.576135 & 2.62993 & 0.856796 & 0.0135506 & 0.0210171 \\
 22 & 0.997026 & 0.904527 & 0.00413374 & 0.15503 & 0.00297359 & 0.841997 & 0.0220931 & -0.576451 & 2.63937 & 0.856801 & 0.0135163 & 0.0215442 \\
 23 & 0.996899 & 0.900451 & 0.00302908 & 0.152489 & 0.00310095 & 0.84441 & 0.0219746 & -0.576792 & 2.64889 & 0.856805 & 0.0134824 & 0.0220608 \\
 24 & 0.996772 & 0.896395 & 0.00193035 & 0.150029 & 0.0032277 & 0.846743 & 0.0218567 & -0.577158 & 2.65849 & 0.856808 & 0.0134487 & 0.0225676 \\
 25 & 0.996646 & 0.892359 & 0.000837518 & 0.147645 & 0.00335385 & 0.849001 & 0.0217393 & -0.57755 & 2.66816 & 0.856809 & 0.0134154 & 0.0230652 \\
 26 & 0.996521 & 0.888344 & -0.000249449 & 0.145333 & 0.0034794 & 0.851188 & 0.0216226 & -0.577966 & 2.67791 & 0.85681 & 0.0133823 & 0.0235541 \\
 27 & 0.996396 & 0.884349 & -0.00133058 & 0.143088 & 0.00360437 & 0.853308 & 0.0215065 & -0.578408 & 2.68774 & 0.856809 & 0.0133495 & 0.0240348 \\
 28 & 0.996271 & 0.880374 & -0.00240591 & 0.140907 & 0.00372877 & 0.855365 & 0.021391 & -0.578876 & 2.69765 & 0.856807 & 0.013317 & 0.0245079 \\
 29 & 0.996147 & 0.876419 & -0.00347546 & 0.138786 & 0.0038526 & 0.857361 & 0.0212761 & -0.579369 & 2.70764 & 0.856804 & 0.0132848 & 0.0249737 \\
 30 & 0.996024 & 0.872484 & -0.00453926 & 0.136723 & 0.00397586 & 0.859301 & 0.0211617 & -0.579887 & 2.7177 & 0.856799 & 0.0132529 & 0.0254326 \\
 31 & 0.995901 & 0.868569 & -0.00559735 & 0.134714 & 0.00409858 & 0.861187 & 0.021048 & -0.580431 & 2.72785 & 0.856794 & 0.0132212 & 0.025885 \\
 32 & 0.995779 & 0.864674 & -0.00664975 & 0.132758 & 0.00422075 & 0.863022 & 0.0209348 & -0.581 & 2.73807 & 0.856788 & 0.0131898 & 0.0263312 \\
 33 & 0.995658 & 0.860799 & -0.00769649 & 0.130851 & 0.00434238 & 0.864807 & 0.0208222 & -0.581596 & 2.74838 & 0.85678 & 0.0131587 & 0.0267715 \\
 34 & 0.995537 & 0.856943 & -0.0087376 & 0.128991 & 0.00446348 & 0.866546 & 0.0207101 & -0.582217 & 2.75877 & 0.856772 & 0.0131279 & 0.0272062 \\
 \textcolor{red}{35} & \textcolor{red}{0.995416} & \textcolor{red}{0.853107} & \textcolor{red}{-0.0097731} & \textcolor{red}{0.127177} & \textcolor{red}{0.00458406} & \textcolor{red}{0.868239} & \textcolor{red}{0.0205987} & \textcolor{red}{-0.582864} & \textcolor{red}{2.76923} & \textcolor{red}{0.856762} & \textcolor{red}{0.0130973} & \textcolor{red}{0.0276355} \\
\bottomrule
\end{tabular}
}% end resizebox
\caption{As in Table~\ref{PQ0:0.2q1:-30}, but with $\hat{Q}_0=0.03$ and captured charge
$\hat{q}_1=-5$ [Fig.~\ref{fig:charge-reversal}, row~B]. The dimensionless charge
$\hat{Q}_n$ again decreases monotonically, crosses the neutral state, and reverses sign,
confirming that the charge-sign reversal is generic rather than tied to a particular
choice of the initial charge. Remaining parameters: $\hat{q}_0=0.2$, $\hat{E}_0=1$,
$\hat{r}=1.9$, $\nu=0.78345$, $\hat{p}_{\phi1}=-19.434$, $\mu_0=0.01\,M_0$, and $M_0=1$.}
\label{PQ0:0.03q1:-5}
\end{table}

\end{document}